\documentclass[aps,preprint,showpacs,preprintnumbers,amsmath,amssymb]{article}
\usepackage{float}
\usepackage{graphicx}
\RequirePackage{graphicx}
\RequirePackage{flushend}
\RequirePackage[colorlinks,citecolor=blue,urlcolor=blue,linkcolor=blue]{hyperref}
\usepackage{slashed}
\usepackage{verbatim}
\usepackage{graphicx}
\usepackage{combelow} 
\usepackage{mathtools}
\usepackage[usenames,dvipsnames,svgnames,table]{xcolor}
\usepackage{soul}
\usepackage{ulem}
\usepackage{braket}
\usepackage{bm}
\usepackage{amssymb} 
\usepackage{authblk}
\usepackage{subcaption,lipsum}

\title{Early Universe production of $W$ bosons in neutrino decays}
\author[1]{Amalia Dariana Fodor \thanks{amalia.fodor97@e-uvt.ro}}
\author[1]{Andru Mihai Buga \thanks{andru.buga02@e-uvt.ro}}
\author[1]{Cosmin Crucean \thanks{cosmin.crucean@e-uvt.ro}}

\affil[1]{Faculty of Physics, West University of Timi\c soara, V. Parvan Avenue 4 RO-300223 Timi\c soara,  Romania}

\begin{document}
\maketitle
\begin{abstract}
In this paper we study, via perturbative methods, the rates of production of $W$ bosons emitted in neutrino decays during the early stages of the Universe. We compute the transition amplitude corresponding to the first order of de Sitter electroweak perturbation theory and study its various limiting cases. The transition rates are derived using minimal subtraction and dimensional regularization. In the end we attempt to obtain the density number of $W$ bosons produced in perturbative transitions in de Sitter spacetime, using the rates obtained in this paper and in previous ones, and analyze the density number with respect to the particle momenta and renormalisation mass $\mu$.
\end{abstract}

\section{Introduction}
The problem of electroweak interactions in curved spacetime \cite{2,WT,cc,43,44,45,46,PTEP} can be approached using the same perturbative formalism as in flat spacetime theory \cite{PR1,PR2,PR3,w2,3,4,5,6,7,8,9,10,11,cr,12,19,20}. The difference is that spacetime expansion, if we refer to the de Sitter Universe, modifies the time modulation of the free modes \cite{2,17,18,PT,PC,22}. This has dramatic physical and computational consequences. But one of the most important features of the theory in the de Sitter geometry is related to the fact that the loss of translational invariance with respect to time allows one to study the first order amplitudes of transitions which are forbidden in flat spacetime theory \cite{15,LL,LL1,23,rc,24,25,26,27,28,29,30,31,b1,b2,cc,cpc,43,44,45,46,PTEP}. These amplitudes count for the generation of fermions and bosons from vacuum, as well as for the emission of bosons by fermions \cite{23,rc,24,25,26,27,28,29,30,31,b1,b2,cc,cpc,43,44,45,46,PTEP}. In the present paper we want to explore these kinds of transitions and establish the density number of particles.

When referring to quantum field theory in curved spacetime, specifically in a Robertson-Walker background, we must mention the important results which concern the free Dirac and Proca field equations, together with their modes \cite{2,22,CML,32,33,34,35}. These results were completed by the analysis of the Dirac and Proca propagators, in both the coordinate and the momentum representations \cite{PT,WT,CRR,COT,rfv,rfv1,38,40,41}. These works established a turning point for the study of fields with spin in curved spacetime, and have given the first insights into how a perturbative field theory could be constructed in the de Sitter geometry by following the methods used in flat spacetime field theory \cite{12,19}. The mechanisms that were involved in the problem of fermion and boson production in the Early Universe were also intensively studied. Important results, which give the first ideas about this subject, can be found in works by E. Schr\"odinger \cite{32} and L. Parker \cite{33} - \cite{35}. Further studies on the subject of particle production were also done, using both perturbative and non-perturbative methods \cite{33} - \cite{35}. However, the results obtained so far do not give a clear picture of the phenomenon of particle production in strong gravitational fields. 

Still, the main conclusion that can be drawn is that particle production phenomena should be placed in the early stages of the universe \cite{w1,24,32,33,34,35,43,44}. In many cases of interest the density number of particles can only be approximated, since the analytical results are divergent. For that reason, methods for removing these divergences need to be employed. In this paper we will refer only to perturbative results, which can be obtained using the methods of flat spacetime theory, adapted to a curved background. This paper is based on results which can be found in \cite{43} - \cite{45}, \cite{PTEP}, where the decay rates for perturbative processes were obtained in conditions of large spacetime expansion, by using a combination of renormalization methods. 

In the present study we want to determine the density number of $W^{\pm}$ bosons, as a result of production in all perturbative first order processes of electroweak theory, where we include their emission by neutrinos and anti-neutrinos in perturbative processes that could take place in the Early Universe. For our study we will use the expanding part of the de Sitter geometry \cite{1}, known as the Poincar\'e patch. The previous two studies \cite{44} and \cite{PTEP}, which discuss the perturbative production of $W^{\pm}$, offer the transition rates in the limit of large expansion as finite quantities. In this limit we obtain the density number of $W^{\pm}$ bosons, and study the evolution of this density number with respect to the particle momenta. 

The paper is organized as follows: in the second section we obtain the amplitude for neutrinos emitting a $W^+$ boson and an electron $\nu \rightarrow W^{+} +e^{-}$. In the third section we establish the partial transition rate as a function dependent on the ratio between the particle masses and the Hubble parameter. The total transition rate is also computed in section three, by using the dimensional regularization method, combined with minimal subtraction. In section four we obtain the density number of $W^{\pm}$ charged bosons that could result in perturbative first order processes, and provide a graphical analysis in terms of the particle momenta and renormalisation parameter $\mu$. Our conclusions are summarized in section five. \hyperref[AppA]{Appendix A} contains the free Dirac and Proca field solutions, while \hyperref[AppB]{Appendix B} contains useful mathematical formulas. In \hyperref[AppC]{Appendix C} we give the result of the transition amplitude computed with longitudinal modes, and explain their contribution to the process $\nu \rightarrow W^{+} +e^{-}$.

\section{The transition amplitude for transversal Proca modes}

The present work is done in the de Sitter geometry described by the metric \cite{1}:
\begin{equation}\label{metr}
ds^2=dt^2-e^{2\omega t}d\vec{x}^2=\frac{1}{(\omega t_{c})^2}(dt_{c}^2-d\vec{x}^2),
\end{equation}
where in the second equality $t_c$ represents the conformal time. The relation between the conformal time and the proper time is given by $t_{c}=-e^{-\omega t}/\omega$, where $\omega$ is the expansion factor or Hubble parameter ($\omega>0$). Our analysis will be done in the conformal chart $t_{c}\in(-\infty,0)$, which covers the expanding portion of de Sitter spacetime \cite{1}. 

For the line element (\ref{metr}) in the Cartesian gauge, the non-vanishing tetrad components are:
\begin{equation}
e^{0}_{\widehat{0}}=-\omega t_c  ;\,\,\,e^{i}_{\widehat{j}}=-\delta^{i}_{\widehat{j}}\,\omega t_c.
\end{equation}

The first order amplitude corresponding to the process $\nu \rightarrow W^{+} +e^{-}$ in the de Sitter background is:
\begin{align}\label{Amp1}
\mathcal{A}_{[{\nu \rightarrow W^{+} +e^{-} }]} &= \frac{i g}{2\sqrt{2}} \int d^4x \sqrt{-g(x)} \, \bar{u}_{\vec{p}\sigma}(x) \gamma^{\hat{\alpha}} e^{\mu}_{\hat{\alpha}}(1-\gamma^5) u_{\vec{p'}\sigma'}^{0}(x) f_{\vec{P}\lambda,\,\mu}^*(x),
\end{align}
where $u_{\vec{p}\sigma}(x),\,u_{\vec{p'}\sigma'}^{0}(x)$ are the massive and massless Dirac field solutions in de Sitter spacetime, respectively, while $f_{\vec{P}\lambda,\,\mu}(x)$ is the Proca field solution \cite{2}. These modes were deduced in terms of the momentum-helicity basis, and their expressions are given in \hyperref[AppA]{Appendix A}.

For transversal Proca modes, that is when $\lambda=\pm 1$, the temporal component of the Proca field solution vanishes \cite{2}. Therefore we can define the amplitude only using the spatial components of the Proca solution:
\begin{align}\label{Amp1}
\mathcal{A}_{[{\nu \rightarrow W^{+} +e^{-} }]}^{[\lambda=\pm1]} &= \frac{i g}{2\sqrt{2}} \int d^4x \sqrt{-g(x)} \, \bar{u}_{\vec{p}\sigma}(x) \gamma^{\hat{i}} e^{j}_{\hat{i}}(1-\gamma^5) u_{\vec{p' }\sigma'}^{0}(x) f_{\vec{P}\lambda=\pm1,\,j}^*(x).
\end{align}

We use the solutions \eqref{solelectron}, \eqref{solneutrin}, \eqref{procasollambda1} for the Dirac and Proca equations \cite{2,22} from \hyperref[AppA]{Appendix A} and replace them in the above formula, keeping in mind that $\bar{u} = u^{\dag} \gamma^{0}$, and that we use the chiral representation for the Dirac matrices. The form of the solutions allows us to split the temporal and spatial integrals. The spatial integral is the same as the one in the Minkowski theory, and results in the Dirac Delta function which is responsible for momentum conservation in the process. 

We solve the temporal integral in terms of $z=e^{-\omega t}/\omega$. Performing this change of variables leads to:
\begin{align}\label{AmpDecayWz}
\mathcal{A}_{[{\nu \rightarrow W^{+} +e^{-}}]}^{[\lambda=\pm 1]} &= -\frac{g}{4\sqrt{2}}\frac{\pi \sqrt{p} e^{\pi k/2} e^{-\pi K/2}}{(2\pi)^{3/2}}\left(\frac{1}{2}-\sigma^\prime \right)\delta^3(\vec{p^\prime}-\vec{p}-\vec{P}) \nonumber\\
&\times \int z\, dz\, e^{ip^\prime z}H_{\frac{1}{2}+ik}^{(2)}(pz)H_{-iK}^{(2)}(Pz)\xi_{\sigma}^{\dag}(\vec{p})\sigma^{j}\xi_{\sigma^\prime}(\vec{p^\prime})\epsilon_{j}(\vec{P},\lambda=\pm 1),
\end{align}
where the orders of the Hankel functions depend on the masses of the electron and $W$ boson, a well as on the Hubble parameter $\omega$, such that $k=m_e/\omega,\,K=\sqrt{\left(M_W/\omega\right)^{2}-1/4}$.

By using the relations between Hankel functions and Bessel $K, J$ functions \eqref{exptok} - \eqref{hankeltoj} from \hyperref[AppB]{Appendix B} \cite{AS,21}, the temporal integral can be expanded as such:
\begin{align}\label{bessel4}
\int_0^\infty dz z^{3/2}K_{-iK}(iPz) \times & \big[i e^{-\pi k}J_{\frac{1}{2}+ik}(pz)J_{-\frac{1}{2}}(p'z)-e^{-\pi k}J_{\frac{1}{2}+ik}(pz)J_{\frac{1}{2}}(p'z)\nonumber\\ 
&-J_{-\frac{1}{2}-ik}(pz)J_{-\frac{1}{2}}(p'z)-iJ_{-\frac{1}{2}-ik}(pz)J_{\frac{1}{2}}(p'z)\big].
\end{align}
For a more detailed explanation of this computation, please see Refs. \cite{44} and \cite{PTEP}.

These transformations lead to the following expression for the transition amplitude:
\begin{align}
\mathcal{A}_{[{\nu \rightarrow W^{+} +e^{-}}]}^{[\lambda=\pm 1]} &= \frac{g\sqrt{i}}{4}\frac{\sqrt{\pi}\sqrt{p p^\prime}e^{\pi k/2}e^{-i \pi/4}}{(2\pi)^{3/2} \,i \cosh{(\pi k)}} \left(\frac{1}{2}-\sigma\prime\right)\delta^3(\vec{p}-\vec{p\prime}-\vec{P})\nonumber\\
&\times \left[T_1 + T_2 + T_3 + T_4\right]\xi_{\sigma}^{\dag}(\vec{p})\sigma^{j}\xi_{\sigma^\prime}(\vec{p^\prime})\epsilon_{j}(\vec{P},\lambda=\pm 1),
\end{align}
where terms $T_1,T_2,T_3,T_4$ represent the four integrals with Bessel functions from expression \eqref{bessel4}. 

The integrals can be solved by using formula \eqref{appell} from \hyperref[AppB]{Appendix B} \cite{AS,21}:
\begin{align}\label{t1}
T_1 &= \frac{i e^{-\pi k}\sqrt{2} p\prime^{-\frac{1}{2}} p^{\frac{1}{2}+ik} (-iP)^{-\frac{5}{2}-ik}}{\Gamma\left(\frac{1}{2}\right)\Gamma\left(\frac{3}{2}+ik\right)} \nonumber \\
&\times \Gamma\left(\frac{\frac{5}{2}+i(k-K)}{2}\right)  \Gamma\left(\frac{\frac{5}{2}+i(k+K)}{2}\right)\nonumber \\
&\times F_{4}\Bigg( \frac{\frac{5}{2}+i(k-K)}{2}, \frac{\frac{5}{2}+i(k+K)}{2}, \frac{1}{2}, \frac{3}{2} +ik,\frac{p\prime^2}{P^2},\frac{p^2}{P^2}\Bigg),
\end{align}

\begin{align}\label{t2}
T_2 &= \frac{e^{-\pi k}\sqrt{2} p\prime^{\frac{1}{2}} p^{\frac{1}{2}+ik} (-iP)^{-\frac{7}{2}-ik}}{\Gamma\left(\frac{3}{2}\right)\Gamma\left(\frac{3}{2}+ik\right)} \nonumber \\
&\times \Gamma\left(\frac{\frac{7}{2}+i(k-K)}{2}\right)  \Gamma\left(\frac{\frac{7}{2}+i(k+K)}{2}\right)\nonumber \\
&\times F_{4}\Bigg(\frac{\frac{7}{2}+i(k-K)}{2}, \frac{\frac{7}{2}+i(k+K)}{2}, \frac{3}{2}, \frac{3}{2} + ik, \frac{p\prime^2}{P^2},\frac{p^2}{P^2}\Bigg),
\end{align}

\begin{align}\label{t3}
T_3 &= -\frac{\sqrt{2} p\prime^{-\frac{1}{2}} p^{-\frac{1}{2}-ik} (-iP)^{-\frac{3}{2}+ik}}{\Gamma\left(\frac{1}{2}\right)\Gamma\left(\frac{1}{2}-ik\right)} \nonumber \\
&\times \Gamma\left(\frac{\frac{3}{2}-i(k+K)}{2}\right)  \Gamma\left(\frac{\frac{3}{2}-i(k-K)}{2}\right)\nonumber \\
&\times F_{4}\Bigg(\frac{\frac{3}{2}-i(k+K)}{2}, \frac{\frac{3}{2}-i(k-K)}{2}, \frac{1}{2}, \frac{1}{2} - ik,\frac{p\prime^2}{P^2},\frac{p^2}{P^2}\Bigg),
\end{align}

\begin{align}\label{t4}
T_4 &= \frac{i^{-1} \sqrt{2} p\prime^{\frac{1}{2}} p^{-\frac{1}{2}-ik} (-iP)^{-\frac{3}{2}+ik}}{\Gamma\left(\frac{3}{2}\right)\Gamma\left(\frac{1}{2}-ik\right)} \nonumber \\
&\times \Gamma\left(\frac{\frac{5}{2}-i(k+K)}{2}\right)  \Gamma\left(\frac{\frac{5}{2}-i(k-K)}{2}\right)\nonumber \\
&\times F_{4}\Bigg(\frac{\frac{5}{2}-i(k+K)}{2}, \frac{\frac{5}{2}-i(k-K)}{2}, \frac{3}{2}, \frac{1}{2} - ik,\frac{p\prime^2}{P^2},\frac{p^2}{P^2}\Bigg),
\end{align}
where $F_4$ is the Appell hypergeometric function \eqref{F4series}. 

Terms $T_1 -T_4$ contain the dependence on the Hubble constant through parameters $k$ and $K$. They also depend on the momenta of the particles involved in the process. As can be observed in the above equations, the de Sitter amplitude has a very complicated mathematical expression, and needs to be analyzed graphically in order to extract physical consequences. In this regard, it is useful to obtain different limits for de Sitter amplitudes, such as the Minkowski limit or the Early Universe limit, when the expansion parameter is much larger than the particles masses. This will be the topic of the next section, where we will obtain the transition rate for the neutrino decay and compute the large expansion limit.

\section{The transition rate}

In this section we want to obtain the dependence of the partial transition rate on the ratio between the particles masses and Hubble constant. For this purpose we rewrite the amplitude in the following general form
\begin{equation}\label{dsrt}
A_{if}=\delta^3(\vec p\,'-\vec p-\vec P)M_{if}I_{if},
\end{equation}
where the temporal integral is denoted by $I_{if}$ and can be written as
\begin{equation}
I_{if}=\int_0^{\infty} dt \mathcal{K}_{if},
\end{equation}
with $\mathcal{K}_{if}$ being the integrand of the temporal integral. In addition, $M_{if}$ represents the the rest of the factors present in amplitude \eqref{AmpDecayWz}.

The transition rate per unit volume in the conformal chart is defined as follows:
\begin{align}\label{rt}
  R_{if}(\omega) &= \frac{1}{2} \sum_{\sigma'\lambda}\frac{1}{(2\pi)^3}\,\delta^3(\vec{p}\,'-\vec{p}-\vec{P}\,)| M_{if}|^2 |I_{if}| \lim_{t\rightarrow \infty}| e^{\omega t}\mathcal{K}_{if}|.
\end{align}

An extended discussion about the definition of the transition rate in de Sitter Universe can be found in \cite{44,PTEP}. We also mention that the first general definition of this rate was given in \cite{cpc}, where the analogy with the definition of the transition rate in Minkowski spacetime was also discussed.

In order to write down the quantities necessary to define the transition rate, we rewrite the amplitude in terms of Hankel and Bessel $K$ functions:
\begin{align}
\mathcal{A}_{[ \nu \rightarrow W^{+} + e^-]}^{[\lambda=\pm 1]} &= \frac{i g e^{\pi k/2}}{\pi\sqrt{2}}\frac{ \sqrt{p}  }{(2\pi)^{3/2}}\left(\frac{1}{2}-\sigma' \right)\delta^3(\vec{p}\,'-\vec{p}-\vec{P})\frac{2i}{\pi} \nonumber\\
&\times \int z\, dz\, e^{ip' z}H^{(2)}_{\frac{1}{2}+ik}(pz)K_{-iK}(iPz)\nonumber\\
&\times\xi_{\sigma}^{\dag}(\vec{p})\sigma^{j}\xi_{\sigma'}(\vec{p}\,')\epsilon_{j}(\vec{P},\lambda=\pm 1).
\end{align}

The transition rate is relevant when $\omega\gg M_{W}$ and $\omega\gg m_{e}$. Thus we can further approximate the index of the modified Bessel function $K_{-iK}(iPz)$ such that \cite{AS,21}:
\begin{equation}
-i\sqrt{\left(\frac{M_{W}}{\omega}\right)^{2}-\frac{1}{4}} = \frac{1}{2}\sqrt{1 - 4\left(\frac{M_{W}}{\omega}\right)^{2}}\approx \frac{1}{2}.
\end{equation}

An analytical result for the transition rate can be obtained if the Bessel $K$ functions are expanded for small arguments using $K_{\nu} (z) \simeq 2^{\nu -1} \Gamma(\nu)/z^\nu$, since for $t\rightarrow \infty$ the argument $z\sim e^{-\omega t}\rightarrow 0$ \cite{AS,21}. This means:
\begin{equation}
  K_{1/2} \left(\frac{Pe^{-\omega t}}{\omega}\right) \simeq \frac{\Gamma(1/2)}{\sqrt{2i\frac{Pe^{-\omega t}}{\omega}}}.
\end{equation}

Hankel functions can also be approximated for small arguments using $H^{(2)}_{\nu}(z)\simeq i 2^{\nu} \Gamma(\nu)/\pi z^\nu$, which in our case translates to the following equation \cite{AS,21}:
\begin{equation}
   H^{(2)}_{1/2+ik} \left(\frac{pe^{-\omega t}}{\omega}\right)\simeq \frac{-i\Gamma(1/2+ik)}{\pi}\left(\frac{2}{\frac{pe^{-\omega t}}{\omega}}\right)^{1/2+ik}.
\end{equation}

Then the limit of the integrand can be computed by using the above approximations, which still preserve the dependence on parameters $k$ and $K$. The result of this limit is:
\begin{align}\label{lim}
  & \lim_{t \rightarrow t_{\infty}} \big| e^{\omega t} \mathcal{K}_{if} \big| = \frac{1}{ \sqrt{p P \cosh(\pi k)}} .
\end{align}

With \eqref{lim} we can write down the final expression for the transition rate:
\begin{align}
   R_{if} &= \frac{1}{2}\sum_{\lambda \sigma} \frac{g^2\pi^2pp'}{32(2\pi)^{3}}\delta^{3}(\vec{p^{\prime}}-\vec{p} - \vec{P}) \frac{|I_{if}| }{\sqrt{Pp\cosh(\pi k)}} \nonumber\\
   &\times |\xi_{\sigma}^{\dag}(\vec{p})\sigma^{i}\xi_{\sigma'}(\vec{p}\,'){\epsilon}^{*}_{i}(\vec{P},\lambda= \pm1)|^{2} 
\end{align}
where $|I_{if}| = \sqrt{I_{if}I_{if}^{*}}$.

The result for the temporal integral is:
\begin{equation}
I_{if}=\frac{e^{i\pi/4}e^{\pi k/2}}{i\sqrt{p'}\cosh(\pi k)}(T_1+T_2+T_3+T_4),
\end{equation}
where $T_1,T_2,T_3,T_4$ are defined in equations (\ref{t1}),(\ref{t2}),(\ref{t3}),(\ref{t4}).

For the sake of examining the behavior of the transition rate, it is useful to perform a graphical analysis of the terms which depend on the ratios of the particle masses and the Hubble parameter. We can group all relevant terms into a single quantity:
\begin{align}\label{Rgraff}
	\mathcal{R} &= \frac{\sqrt{p p'} e^{\pi k/2}}{\cosh{(\pi k)}}  \sqrt{\frac{(T_1 + T_2 + T_3 + T_4)(T_1 + T_2 + T_3 + T_4)^{*}}{P \cosh{(\pi k)}}}.
\end{align}

\begin{figure}[h!]
        \centering
	\includegraphics[width=7.5cm]{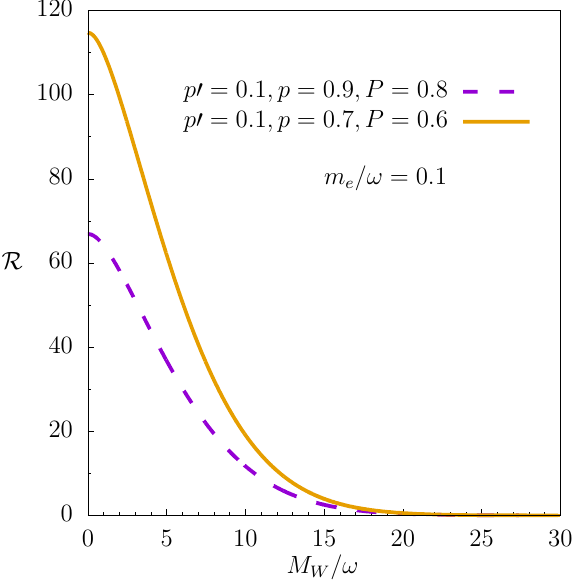}
	\caption{Transition rate $\mathcal{R}$ \eqref{Rgraff} as a function of $M_W/\omega$, for given values of the momenta and a fixed value of $m_e/\omega$.}
	\label{fig.1}
\end{figure}

\begin{figure}[h!]
	\centering
	\includegraphics[width=7.5cm]{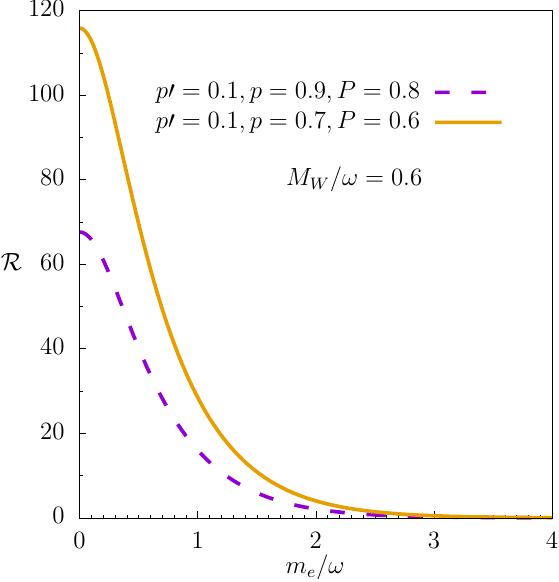}
	\caption{Transition rate $\mathcal{R}$ \eqref{Rgraff} as a function of $m_e/\omega$, for given values of the momenta and a fixed value of $M_W/\omega$.}
	\label{fig.2}
\end{figure}

According to Figures \ref{fig.1} and \ref{fig.2}, the transition rate reaches a maximum for small values of the ratio between the particle masses and the Hubble constant. As the ratio between the masses and the Hubble constant increases, the transition rate tends to zero. This confirms that the process of $W^{\pm}$ production by neutrinos is no longer allowed in the Minkowski limit, when $\omega \rightarrow 0$. The vanishing rate in the Minkowski limit is also confirmed by the fact that the probability density of this process also vanishes in this limit, since the transition rate is proportional with the factors that define the square modulus of the amplitude $\mathcal{R} \sim \left|\mathcal{A}_{[{\nu \rightarrow W^{+} +e^{-}}]}\mathcal{A}_{[{\nu \rightarrow W^{+} +e^{-}}]}^* \right|$.


\newpage
\subsection{The transition rate in the limit of large expansion}

The present subsection is dedicated to the computation of the total rate of transition of the process in the limit when $m/\omega=M_W/\omega=0$, which corresponds to the large expansion conditions of the Early Universe. 

In this particular limit, the transition amplitude \eqref{AmpDecayWz} for the process $\nu \rightarrow W^{+} +e^{-}$ depends only on Hankel functions of order $1/2$:
\begin{align}
\mathcal{A}_{[{\nu \rightarrow W^{+} +e^{-}}]}^{[\lambda=\pm1]} &= \frac{ g}{4\sqrt{2}}\frac{\pi\sqrt{p}e^{-i\pi/4}}{(2\pi)^{3/2}} \left(\frac{1}{2}-\sigma'\right)\delta^3(\vec{p}\,'-\vec{p}-\vec{P}) \nonumber\\
&\times\int_{0}^{\infty} dz\, z \,e^{i p\, 'z}H_{1/2}^{(2)} (pz)H_{1/2}^{(1)} (Pz)\xi^{\dag}_{\sigma}(\vec{p})\vec{\sigma}\cdot\vec{\epsilon}(\vec{n}_{\mathcal{P}},\lambda)\xi_{\sigma'}(\vec{p}\,').
\end{align}

This expression is suitable for analytical computations, since the special functions in the general case reduce to simple polynomial functions of the particle momenta. 
In this case the temporal integral can be solved by adding a factor of $\exp(-\epsilon z)$, with $\epsilon >0$, which leads to:
\begin{align}
\mathcal{A}_{[{\nu \rightarrow W^{+} +e^{-}}]}^{[\lambda = \pm 1]} &= \frac{ g}{4\sqrt{2}}\frac{\pi\sqrt{p}e^{-i\pi/4}}{(2\pi)^{3/2}} \left(\frac{1}{2}-\sigma'\right) \delta^3(\vec{p}\,'-\vec{p}-\vec{P}) \,\xi^{\dag}_{\sigma}(\vec{p})\vec{\sigma}\cdot\vec{\epsilon}(\vec{n}_{\mathcal{P}},\lambda)\xi_{\sigma'}(\vec{p}\,')\nonumber\\
&\times\frac{-2i}{\pi\sqrt{pP}(p'-p-P)}
\end{align}
where, if we use the notations from \eqref{dsrt} we find that
\begin{align}\label{ratalimmif}
M_{if}=\frac{ g}{4\sqrt{2}}\frac{\pi\sqrt{p}e^{-i\pi/4}}{(2\pi)^{3/2}} \left(\frac{1}{2}-\sigma'\right)\xi^{\dag}_{\sigma}(\vec{p})\vec{\sigma}\cdot\vec{\epsilon}(\vec{n}_{\mathcal{P}},\lambda)\xi_{\sigma'}(\vec{p}\,'), 
\end{align}
while the result for the temporal integral is:
\begin{align}\label{ratalimiif}
    I_{if} = \frac{-2i}{\pi\sqrt{pP}(p'-p-P)}.
\end{align}

The factor $|M_{if} |^2$ has to be summed over the final polarizations of the particles. For computing this sum we will consider that the particles in the final state are emitted in the same direction as the neutrino in the initial state. The neutrino and $W$ boson will have the same orientation, while the electron will have opposite orientation, i.e. $\vec P=-P \vec e_3;\,\vec p=p \vec e_3;\, \vec p\,'=-p' \vec e_3$. In this case the bispinor summation gives only one non-vanishing term, the one corresponding to $\sigma' =\sigma= -1/2$:
\begin{equation}
    \xi^{\dag}_{-1/2}(\vec{p})\vec{\sigma}\cdot\vec{\epsilon}_{-1}(\vec P,\lambda)\xi_{-1/2}(\vec{p}\,')=\sqrt{2}  .
\end{equation}

Next we replace \eqref{ratalimmif} and \eqref{ratalimiif}, along with the large time limit of the integrand present in the temporal integral:
\begin{align}
    \lim_{t\rightarrow \infty} |e^{\omega t} K_{if}|&= \lim_{t\rightarrow \infty} \biggl|e^{\omega t} e^{-\omega t} e^{i(p'-P-p)\frac{e^{-\omega t}}{\omega}}\frac{i\pi}{\sqrt{Pp}}\biggl|\nonumber\\
    &=\frac{\pi}{\sqrt{Pp}},
\end{align}
in the transition rate formula defined earlier \eqref{rt}.

Consequently we reach the final form of the transition rate in the large expansion limit:
\begin{equation}
 R_{if}=\frac{g^2\pi}{8(2\pi)^6}  \frac{\delta^3(\vec{p}\,'-\vec{p}-\vec{P}) }{P(p'-p-P)}  .
\end{equation}

The total transition rate in the limit of large expansion is obtained by integrating after the final momenta of the electron and $W^+$ boson:
\begin{align}
  R_{\nu\rightarrow W^++ e^-} = \int d^3p \int d^3 P \,R_{if}
\end{align}

One of the momentum integrals can be eliminated by using the Dirac Delta function:
\begin{align}\label{ratatotint}
  I &= \frac{1}{(2\pi)^3} \int d^3p \int d^3P \frac{\delta^3 (\vec{p}\,'-\vec{p} - \vec{P} )}{P(p'-p-P)} \nonumber\\
    &= \frac{1}{(2\pi)^3} \int d^3P \frac{1}{2P(p'-P)},
\end{align}
where the above result is obtained by taking the momenta orientations discussed earlier in this section. 

The remaining integral is divergent, which means we have to use regularization methods \cite{WF,GHV,IT,GT,PV,HGF} to obtain a finite result. We start with dimensional regularization \cite{GHV,GT}, and rewrite integral \eqref{ratatotint} in the general case of $D$ dimensions:
\begin{align}\label{id4}
  I(D) = \frac{2\pi^{D/2}}{2P (2\pi)^D \Gamma(D/2)} \int_{0}^{\infty} dP \frac{P^{D-2}}{p'-P}\,.
\end{align}

By making the substitution $P=-p'y$, we arrive at the integral which defines the Beta Euler function:
\begin{align}\label{id5}
  I(D) &= \frac{2\pi^{D/2}(-p')^{D-2}}{2(2\pi)^D \Gamma(D/2)} \int_{0}^{\infty} \frac{y^{D-1}}{1+y} dy \nonumber\\
  &= \frac{2\pi^{D/2}(-p')^{D-2} \Gamma(D-1) \Gamma(4-D)}{(2\pi)^D 2 \Gamma(D/2)(2-D)(3-D)}.
\end{align}

The above result is still divergent for $D=3$, due to the factor $(3-D)^{-1}$. A way for solving this divergence is to apply the minimal subtraction method \cite{GT}. Firstly we compute the residue of the integral in $D=3$:
\begin{align}
  \text{Res} \, I(D)= \lim_{D\rightarrow 3} (3-D) I(D) = -\frac{p'}{4\pi^2}.  
\end{align}

Afterwards we can define the regularized integral as:
\begin{align}\label{ratatotreg}
  I(D)_r &= I(D) - \frac{p'\mu^{D-3}}{4\pi^2 (3-D)}\nonumber \\
  &= \frac{1}{3-D} \bigg( \frac{2\pi^{D/2}(p')^{D-2} \Gamma(D-1) \Gamma(4-D)}{2P(2\pi)^D (2-D) \Gamma(D/2)} + \frac{p'\mu^{D-3}}{4\pi^2} \bigg),
\end{align}
where the counter-term
\begin{equation}
  \frac{\mu^{s}\,\text{Res}\, I(D)}{3-D}  
\end{equation}
with $s=D-3$ and mass parameter $\mu$ has been chosen such that it has the same dimension as $ I(D)$. 

The expansion around the value $D=3$ of the paranthesis in \eqref{ratatotreg} will result in terms that cancel out the divergent factor $(3-D)^{-1}$:
\begin{align}\label{id5}
I(D)_r = \,\frac{1}{3-D} \times \Bigg\{ \frac{p'(3-D)}{8\pi^2} \bigg[2-\gamma+ \ln \Big( \frac{\pi\mu^2}{p'^2} \Big) \bigg] + \mathcal{O}\big((D-3)^2\big)\Bigg\}.
\end{align}

It is important to point out that the same result for the regularized integral can be obtained from equation \eqref{ratatotint} if we solve first for the integral over $P$, and then the one over $p$.

The term in front of parenthesis will simplify the factor $D-3$ of the first term, while terms proportional with $(D-3)^2$ will vanish when taking the limit $D=3$. The transition rate is finally:
\begin{align}\label{ratatotlimneu}
  R_{\nu \rightarrow W^++e^-}
  =\frac{ G_FM_W^2\, p'}{8\sqrt{2}\pi^2(2\pi)^3 } \bigg[2-\gamma+ \ln \Big( \frac{\pi\mu^2}{p'^2} \Big) \bigg].
\end{align}

Relation \eqref{ratatotlimneu} represents the total decay rate for the process $\nu \rightarrow W^{+} +e^{-}$ in the limit of large expansion, which corresponds to the Early Universe. A similar derivation can be done for the transition rate of the antineutrino decay $\widetilde{\nu} \rightarrow W^{-} +e^{+}$. The outcome is that the total rates for both of these transitions are equal in this limit:
\begin{equation}
R_{\nu \rightarrow W^++e^-} =R_{\widetilde{\nu} \rightarrow W^-+e^+}.
\end{equation}

\section{Density number of particles}

In this section we will demonstrate how to obtain the density number of particles from the transition rates established in the previous section, and also in previous papers \cite{44,PTEP} which discuss other transitions involving $W$ bosons in de Sitter spacetime. Since in the conditions of rapid expansion of the Early Universe the notion of thermal equilibrium is not well defined, we cannot define the density number of $W$ bosons obtained in neutrino decays in  of the density number of neutrinos. This is because the density numbers of neutrinos obtained so far were computed using the distribution function and the hypothesis of thermal equilibrium. For these reasons we will define the density number of $W$ bosons as the ratio between the production rate and the decay rate for the $W$ boson, since this definition establishes a clear picture of the balance between the production and decay of $W$ bosons in the early universe. Then we can use the general equation that gives the variation of the density number in unit volume and in unit time, which has the meaning of the total decay rate per unit volume \cite{gar}:
\begin{equation}\label{numberofW}
\frac{dN}{dt\,dV}=\frac{R_{\text{transition}}}{V} .
\end{equation}

In order to have a clear picture of the number of $W$ bosons that are actually produced, the balance between the production processes and the decay processes must be taken into account. The number of $W$ bosons produced in perturbtive electroweak interactions in the Early Universe can be computed by  substracting the total decay rate from the total production rate, both computed in the de Sitter metric by taking the limit $\text{mass}/\omega\rightarrow 0$. Our proposal is to define the number of bosons in a time interval and in a volume interval, since the background is expanding. For that we use equation \eqref{numberofW}:
\begin{align}
    \frac{dN}{dt\,dV}=\frac{R_{\text{production}}(W)-R_{\text{decay}}(W)}{V}
\end{align} 
where, $R_{\text{production}}(W)$ is the total rate of production of $W$ bosons in perturbative processes in de Sitter spacetime, while $R_{\text{decay}}(W)$ is the total decay rate for the $W$ bosons in de Sitter spacetime.

The total rate of production is obtained by adding the first order electroweak transitions in which $W$ bosons are produced in the de Sitter universe. That includes the production of $W$ bosons from vacuum \cite{44} and the production of $W$ bosons in emission processes by fermions \cite{PTEP}, these processes being possible only in the de Sitter metric. The decays for the three families of neutrinos were studied in the present paper, in the previous sections, with the mention that the results for the transition rates are the same for each neutrino family. The same applies to all three leptonic families which can emit $W$ bosons \cite{PTEP}. These processes are written below for the $W^-$ boson:
\begin{align}
   & \tilde{\nu}_{e}\rightarrow W^{-}+ e^{+}   \quad\quad\quad\quad\,\quad  e^{-} \rightarrow W^{-}  + \nu_e \nonumber\\
   & \tilde{\nu}_{\mu}\rightarrow W^{-}+\mu^{+}   \quad\quad\quad\quad\quad  \mu^{-} \rightarrow W^{-} + \nu_\mu \nonumber\\
   &\tilde{\nu}_{\tau} \rightarrow W^{-}+ \tau^{+}   \quad\quad\quad\quad\,\quad  \tau^{-} \rightarrow W^{-} + \nu_\tau, \nonumber
\end{align}
and can be obtained also for $W^+$, the transition rates remaining the same.

To the above production processes one must add the spontaneous production of $W^{\pm}$ bosons from the de Sitter vacuum \cite{44}:
\begin{align}
   &vac \rightarrow W^{-}+e^{+}+\nu_{e}   \nonumber\\
   &vac \rightarrow W^{-}+\mu^{+} + {\nu}_{\mu}  \nonumber\\
   &vac \rightarrow W^{-}+ \tau^{+} +{\nu}_{\tau}  . \nonumber
\end{align}

Summing all these contributions we obtain the total transition rate of production for $W$ bosons:
\begin{align}
  R_{\text{production}}(W^-)=3\left[R_{vac \rightarrow W^{-}+e^{+}+\nu_{e}}+R_{e^{-} \rightarrow W^{-}  + \nu_e}+R_{\tilde{\nu}_{e}\rightarrow W^{-}+ e^{+}}\right],
\end{align}
where the numerical factor 3 stands for all the leptonic possible transitions. We also mention that the same total production rate is obtained for the $W^+$ boson, i.e. $ R_{\text{production}}(W^-)= R_{\text{production}}(W^+)$. 

The transition rates for emission of $W$ bosons from electrons \cite{PTEP} and spontaneous production of $W$ bosons from vacuum \cite{44} are given below: 
\begin{align}
\label{rvac} R_{vac \rightarrow W^{-}+e^{+}+\nu_{e}}&=\frac{\, G_FM_W^2\, }{8\sqrt{2}(2\pi)^3}\frac{M_W}{2}\bigg[\ln \Big( \frac{4\pi\mu^2}{M_W^2} \Big)+ \psi \Big(\frac{3}{2}\Big) \bigg], \\
\label{reltoW} R_{e^{-} \rightarrow W^{-}  + \nu_e}&=\frac{\, G_FM_W^2\,p }{8\sqrt{2}\pi^2(2\pi)^3 }\bigg[2-\gamma+ \ln \Big( \frac{\pi\mu^2}{p^2} \Big) \bigg].
\end{align}

 When computing the total decay rate for the $W$ bosons, the decay into lepton-antilepton pairs and quark-antiquark pairs must be taken into account. In the Minkowski field theory the total decay rate of the $W$ boson is $R_d=2.5$ GeV \cite{rat}, with the observation that the decay into quarks accounts for $ 60$ \% of the total decay rate \cite{rat}. 

The total decay rate for the $W$ boson will comprise of two terms, one that will take into account the leptonic decays $R_{W^\rightarrow l+\widetilde{l}}$, and one that will take into account the decays into quark-antiquark pairs $R_{W\rightarrow q+\bar{q}}$ \cite{rat}. 

In the case of leptonic decays, the total transition rate in the conditions of the Early Universe is \cite{PTEP}: 
\begin{align}\label{decayW}
  R_{W^-\rightarrow e^-+\widetilde{\nu}}=R_{W^+\rightarrow e^++\nu}
  =\frac{ G_FM_W^2\, P}{4\sqrt{2}\pi^2(2\pi)^3 } \bigg[ \ln \Big( \frac{4\pi\mu^2}{P^2} \Big) + \psi \Big(\frac{3}{2}\Big) \bigg],
\end{align}
with the observation that, in the first order of perturbation theory, the partial widths are equal for all possible decays \cite{rat}:
\begin{align}
   &W^{-} \rightarrow  e^{-} + \tilde{\nu}_{e}  \quad\quad\quad\quad\,\quad  W^{+} \rightarrow e^{+} + \nu_e \nonumber\\
   &W^{-} \rightarrow \mu^{-} + \tilde{\nu}_{\mu}  \quad\quad\quad\quad\quad  W^{+} \rightarrow \mu^{+} + \nu_\mu \nonumber\\
   &W^{-} \rightarrow \tau^{-} + \tilde{\nu}_{\tau}  \quad\quad\quad\quad\,\quad  W^{+} \rightarrow \tau^{+} + \nu_\tau. \nonumber
\end{align}
Our decay rate into leptons must be multiplied by a factor of $3$ if we take into account just one of the charged bosons, and by a factor of $6$ if we take into account both charged bosons.

The decays of $W$ bosons into quark-antiquark pairs in de Sitter spacetime have not been studied so far, but the main step will be to consider the Lagrangian of interaction with quarks, adapted to a curved background. Since in the Minkowski field theory, the $W$ boson decay into quarks-antiquarks $R_{W\rightarrow q+\bar{q}}$ accounts for $60 \,\%$ of the total decay rate, we will consider that the same principle applies in de Sitter spacetime. Therefore we can write:
\begin{equation}
    R_{W\rightarrow q+\bar{q}}=\frac{6}{4}R_{W^\rightarrow l+\widetilde{l}} \,.
\end{equation}

That being the case, we can express the total decay rate of $W^+$ bosons in terms of the $W^+$ rate of decay into leptons:
\begin{align}\label{Wplusdecayrate}
   R_{\text{decay}}(W^-)=  R_{\text{decay}}(W^+)=\frac{10}{4}R_{W^\rightarrow l+\widetilde{l}}=\frac{30}{4}R_{W^+\rightarrow e^++\nu_e}
   =\frac{30}{4}R_{W^-\rightarrow e^-+\widetilde{\nu_e}},
\end{align}
where the last equality was obtained by considering all three leptonic decays of the $W^+$ boson in de Sitter spacetime, $R_{W^+\rightarrow l+\widetilde{l}}=3R_{W^+\rightarrow e^++\nu}$.

Then the density number of $W^-$ bosons in time interval $\Delta t$, and volume interval $\Delta V$, can be estimated with the help of formulas \eqref{numberofW} and \eqref{Wplusdecayrate}:
\begin{align}\label{nt}
     \frac{\Delta N_{W^-}}{\Delta t\,\Delta V}=\frac{3\, G_FM_W^2\, }{8\sqrt{2}\pi^2(2\pi)^3 V}\left\{\frac{\pi^2 M_W}{2}\bigg[\ln \Big( \frac{4\pi\mu^2}{M_W^2} \Big) + \psi \Big(\frac{3}{2}\Big) \bigg] + p\bigg[2-\gamma+ \ln \Big( \frac{\pi\mu^2}{p^2} \Big) \bigg]\right.\nonumber\\
\left.+p'\bigg[2-\gamma+ \ln \Big( \frac{\pi\mu^2}{p'^2} \Big) \bigg]
     -5P\bigg[\ln \Big( \frac{4\pi\mu^2}{P^2} \Big) + \psi \Big(\frac{3}{2}\Big) \bigg]\right\}.
\end{align}

The density number of  $W^-$ bosons must be equal with the density number of $W^+$ bosons, since the decay rates for both charged bosons are equal, while the rates of decay for neutrinos and antineutrinos are also equal $R_{\nu\rightarrow W^++e^-}=R_{\tilde{\nu}\rightarrow W^-+e^+}$. This leads to the total density number of $W$ bosons in time interval $\Delta t$, and volume interval $\Delta V$ being:
\begin{align}\label{nw}
   \frac{\Delta N_{W}}{\Delta t\,\Delta V}=2\frac{\Delta N_{W^-}}{\Delta t\,\Delta V} ,
\end{align}
since there are equal probabilities for spontaneous production of $W^-$ and $W^+$ bosons in various perturbative processes.

The ratio between the density numbers of produced $W$ bosons in perturbative processes and the density number of bosons that decay is  then:
\begin{align}\label{nnd}
   \frac{\Delta N_{W^-}}{\Delta N_{D,W^-}}=  \frac{\frac{\pi^2 M_W}{2}\bigg[\ln \Big( \frac{4\pi\mu^2}{M_W^2} \Big) + \psi \Big(\frac{3}{2}\Big) \bigg]+p\bigg[2-\gamma+ \ln \Big( \frac{\pi\mu^2}{p^2} \Big) \bigg]+p'\bigg[2-\gamma+ \ln \Big( \frac{\pi\mu^2}{p'^2} \Big) \bigg]}{5P\bigg[\ln \Big( \frac{4\pi\mu^2}{P^2} \Big) + \psi \Big(\frac{3}{2}\Big) \bigg]},
\end{align}
where $ N_{D,W^-}$ is the number of $W$ bosons that decay in leptons and quarks. In the above equation the conditions $4\pi\mu^2\geq M_W^2\,; \pi\mu^2\geq p^2\,; \pi\mu^2\geq p'^2$ are mandatory in the logarithms such that we avoid negative values. 

From equations \eqref{nt} and \eqref{nw} it can be seen that the density number of $W$ bosons is a quantity which depends on the particle momenta and the regularization parameter $\mu$. Another consequence is that the density number of produced $W$ bosons begins to drop as the Universe expanded and the decay rate began to increase. This happens when the energy of the background becomes smaller than the rest energy of the $W$ bosons.  This fact can also be seen from a graphical analysis of the density numbers. 

To perform this graphical analysis, we have given values for parameters $p,p',P,\mu$, since the density number of $W$ bosons depends on the initial momenta of the particles $p,p',\,P$,  and on the renormalization mass $\mu$. Usually we will plot with respect to one of these parameters while keeping the other ones fixed (the values given to the fixed parameters can be found on each figure). 

\begin{figure}[H]
        \centering
	\includegraphics[width=10.5cm]{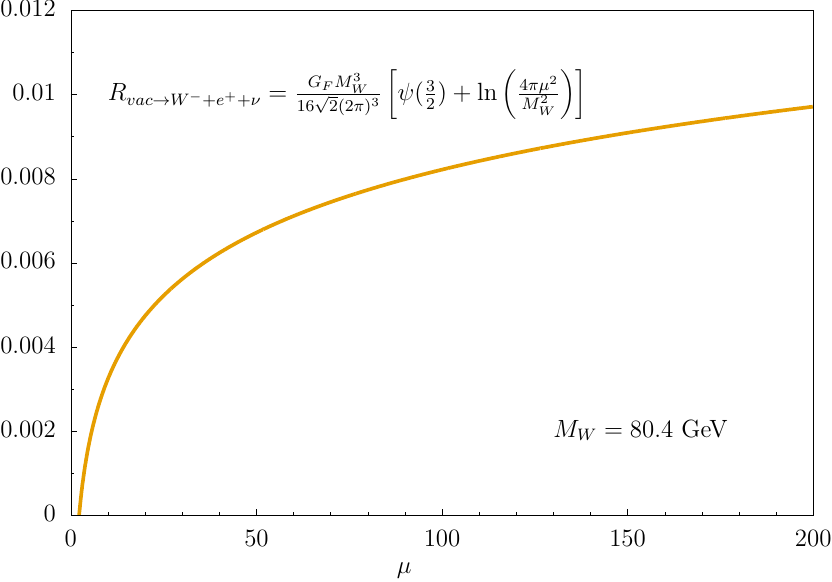}
	\caption{The vacuum transition rate \eqref{rvac}, computed in \cite{44}, as a function of the renormalisation mass $\mu$. }
	\label{fig.3}
\end{figure}

First we plot the transition rates in terms of momenta and parameter $\mu$. In figure \ref{fig.3} we have the vacuum transition rate with respet to $\mu$. We can see that the transition rate increases with $\mu$.

\begin{figure}[H]
        \centering
	\includegraphics[width=10.5cm]{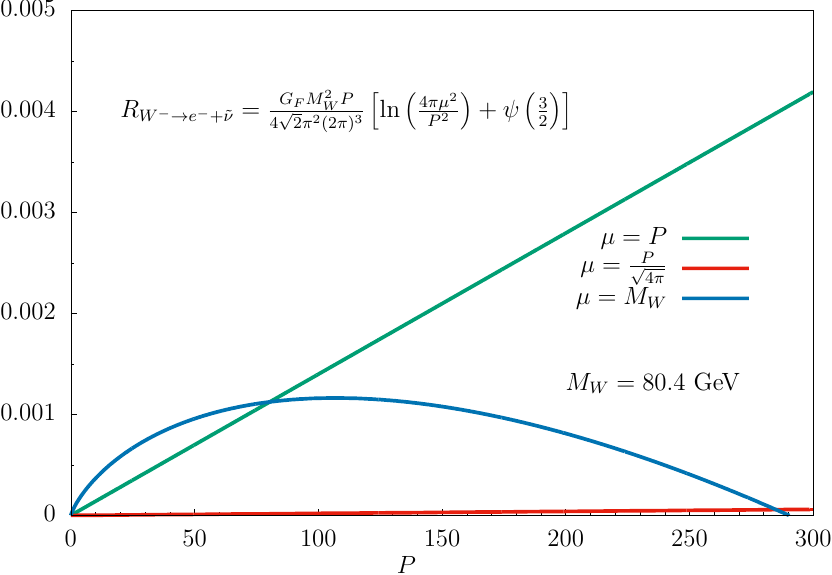}
	\caption{The $W$ decay rate \eqref{decayW}, obtained in \cite{PTEP}, with respect to the $W$ boson momentum $P$, for different values of the renormalisation mass parameter $\mu$.}
	\label{fig.4}
\end{figure}

\begin{figure}[H]
        \centering
	\includegraphics[width=10.5cm]{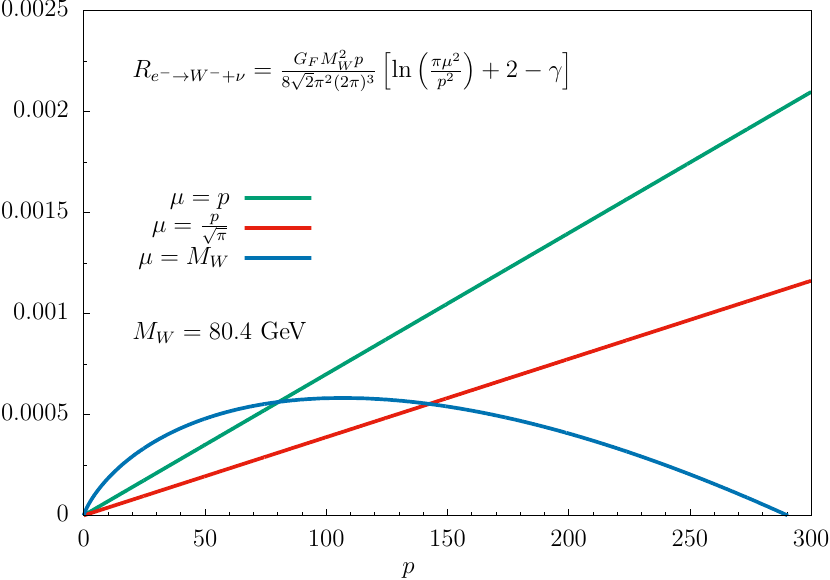}
	\caption{The $W$ boson emission from electron rate \eqref{reltoW}, obtained in \cite{PTEP}, with respect to the electron momentum $p$, for different values of the renormalisation mass parameter $\mu$.}
	\label{fig.5}
\end{figure}

\begin{figure}[H]
        \centering
	\includegraphics[width=10.5cm]{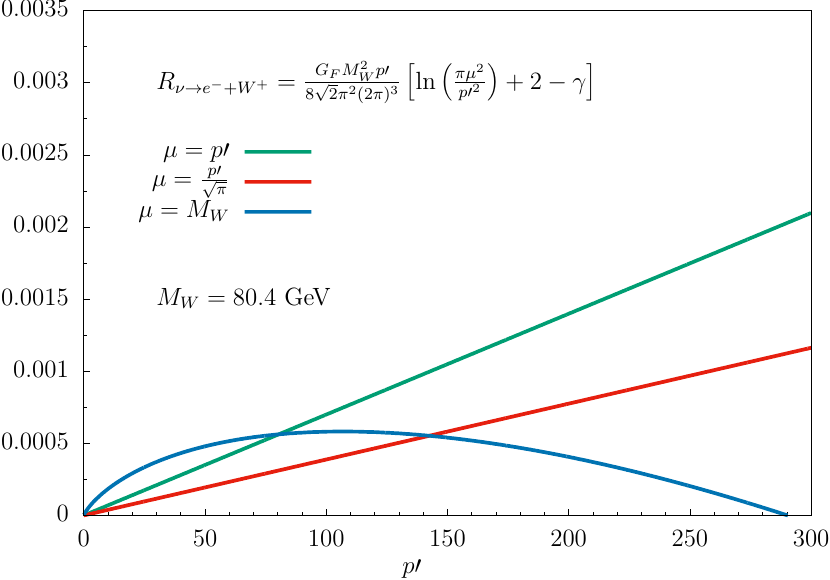}
	\caption{The $W$ boson emission from neutrino rate \eqref{ratatotlimneu}, obtained in the previous section, with respect to the neutrino momentum $p\prime$, for different values of the renormalisation mass parameter $\mu$.}
	\label{fig.6}
\end{figure}

Figures \ref{fig.4} - \ref{fig.6} give the dependence of the total transition rates in terms of the momenta of the particles. The red and green lines give the linear dependence in the case when the logarithm vanishes, and when the logarithm is a positive number, respectively. The blue line represents the situation in which the regularization parameter $\mu$ equals the $W$ boson mass. In this case the argument of the logarithm dictates the graphical behavior, and the transition rates vanish for the cases when the conditions $4\pi\mu^2\geq M_W^2\,; 4\pi\mu^2\geq P^2\,; \pi\mu^2\geq p^2\,; \pi\mu^2\geq p'^2$ are no longer valid. It can be observed that all transition rates increase as the ratio between $\mu$ and the respective particle momentum increases. 

Moreover, one can observe in figure \ref{fig.4} that the leptonic decay rate of the $W$ boson in the limit of large expansion doesn't surpass the Minkowski rate, which for this process is $R_{\,W^- \rightarrow e^- + \tilde{\nu}}^{\,\text{Minkowski}} \simeq 2,25$ GeV, even for momenta as large as $P=300$ GeV. This is consistent with the assumption that the $W$ bosons could exist as stable particles in the Early Universe, when the background energy (due to spacetime expansion) was larger than the rest energy of such heavy particles, and thus decayed less.

\begin{figure}[H]
     \centering
	\includegraphics[width=10cm]{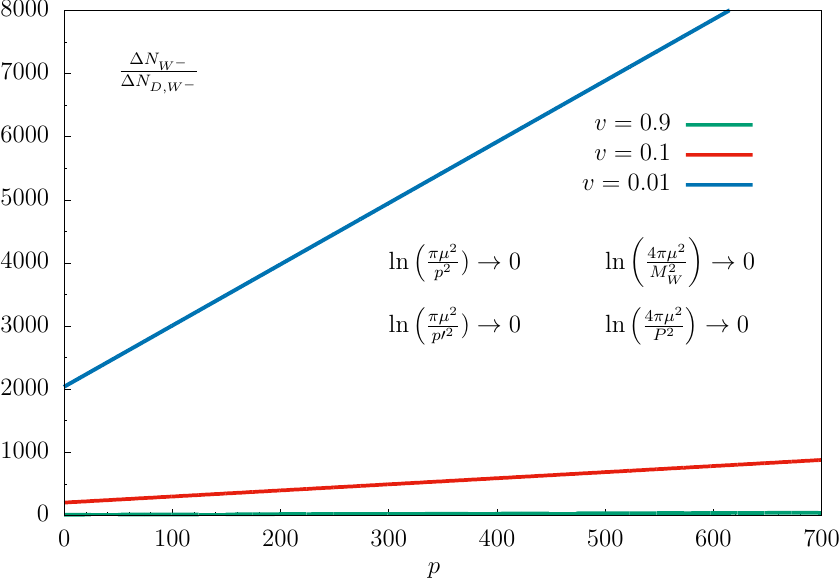}
    \includegraphics[width=10cm]{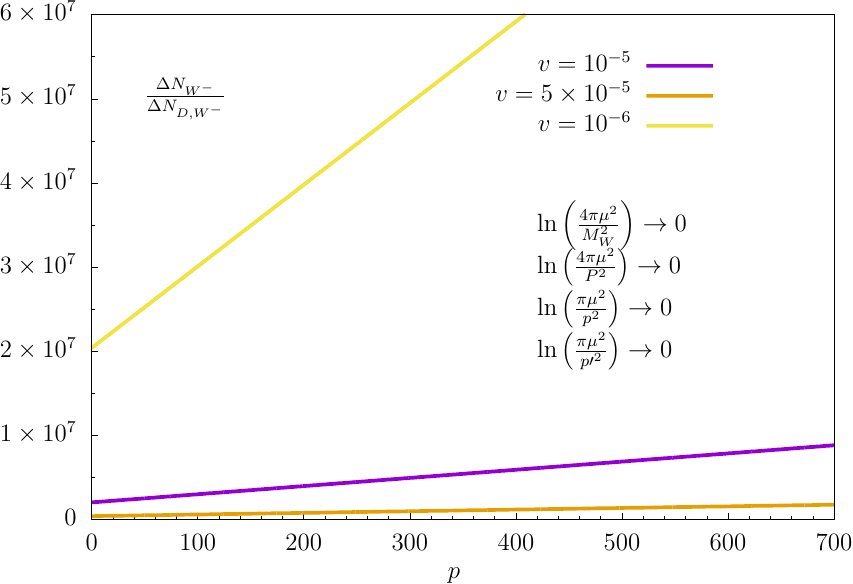}
	\caption{The ratio between the density numbers of produced $W$ bosons in perturbative processes and the density number of decaying $W$ \eqref{nnd} bosons with respect to the electron momentum $p$. Here the $W$ boson momentum is $P= \gamma\, M_{W}\, v$ and the neutrino momentum was taken to be $p\prime = 200$ GeV, where $M_{W}=80.4$ GeV is the $W$ boson mass and $\gamma=1/\sqrt{1-v^2}$ is the Lorentz factor. The ratio was taken at various relativistic (top figure) and norelativistic (bottom figure) speeds (we take $c=1$), for the case when all logarithms in \eqref{nnd} go to 0.}
	\label{fig.8}
\end{figure}

\begin{figure}[H]
     \centering
	\includegraphics[width=10cm]{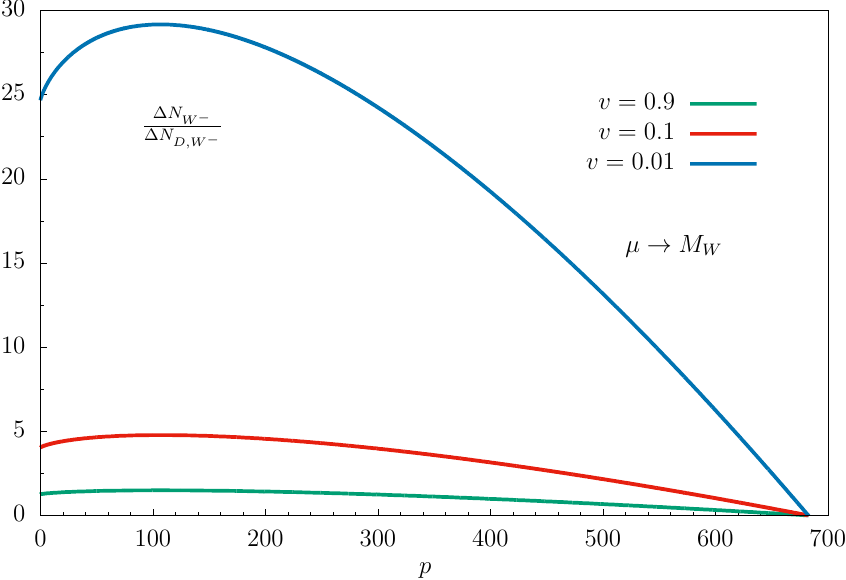}
    \includegraphics[width=10cm]{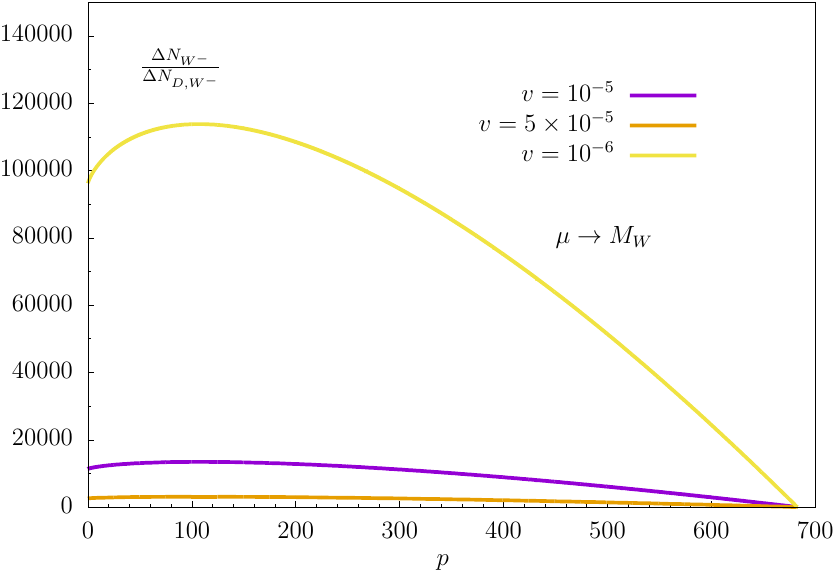}
	\caption{The ratio between the density numbers of produced $W$ bosons in perturbative processes and the density number of decaying $W$ \eqref{nnd} bosons with respect to the electron momentum $p$. Here the $W$ boson momentum is $P= \gamma\, M_{W}\, v$ and the neutrino momentum was taken to be $p\prime = 200$ GeV, where $M_{W}=80.4$ GeV is the $W$ boson mass and $\gamma=1/\sqrt{1-v^2}$ is the Lorentz factor. The ratio was taken at various relativistic (top figure) and norelativistic (bottom figure) speeds (we take $c=1$), for the case when $\mu=M_W$ in \eqref{nnd}.}
	\label{fig.9}
\end{figure}

\begin{figure}[H]
        \centering
	\includegraphics[width=10cm]{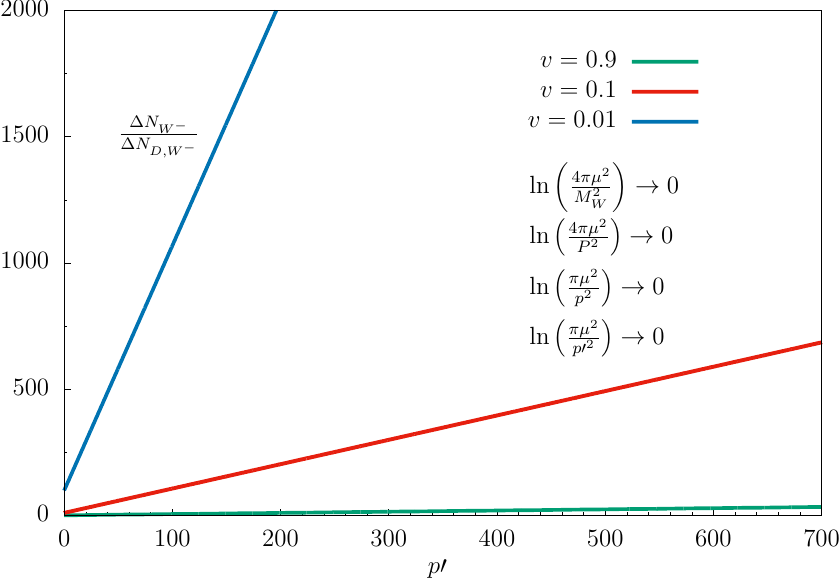}
    \includegraphics[width=10cm]{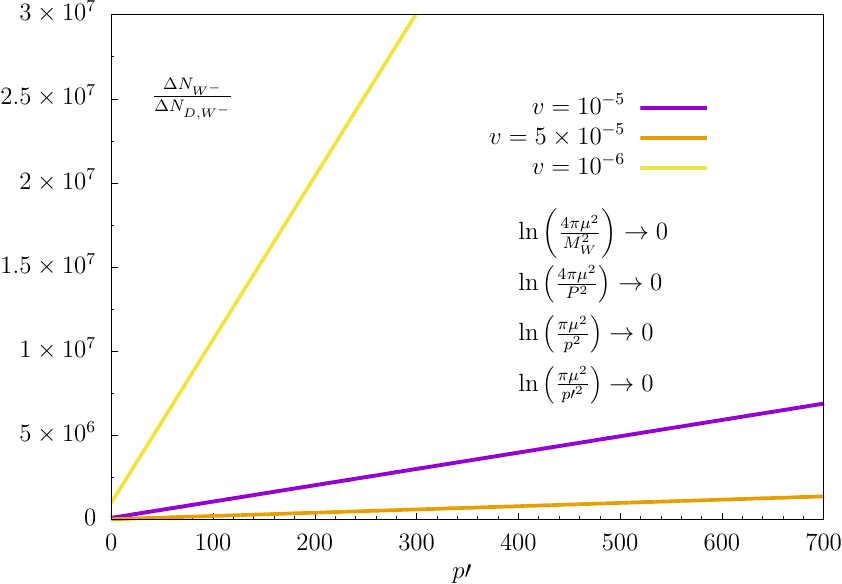}
	\caption{The ratio between the density numbers of produced $W$ bosons in perturbative processes and the density number of decaying $W$ \eqref{nnd} bosons with respect to the neutrino momentum $p\prime$. Here the $W$ boson momentum is $P= \gamma\, M_{W}\, v$ and the electron momentum is $p = \gamma\, m_{\text{el}}\, v$, where $M_{W}=80.4$ GeV is the $W$ boson mass, $m_{\text{el}}=0.511$ MeV and $\gamma=1/\sqrt{1-v^2}$ is the Lorentz factor. The ratio was taken at various relativistic (top figure) and norelativistic (bottom figure) speeds (we take $c=1$), for the case when all logarithms in \eqref{nnd} go to 0.}
	\label{fig.10}
\end{figure}

\begin{figure}[H]
        \centering
	\includegraphics[width=10cm]{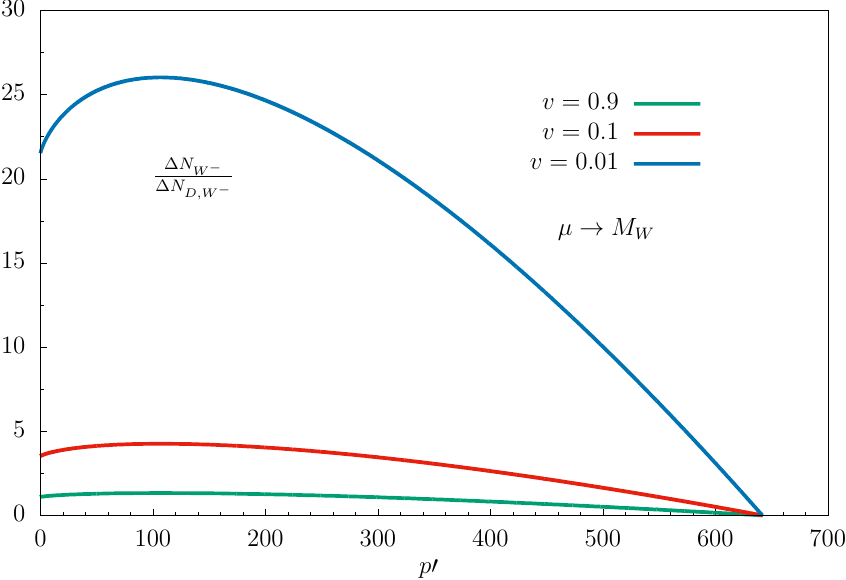}
    \includegraphics[width=10cm]{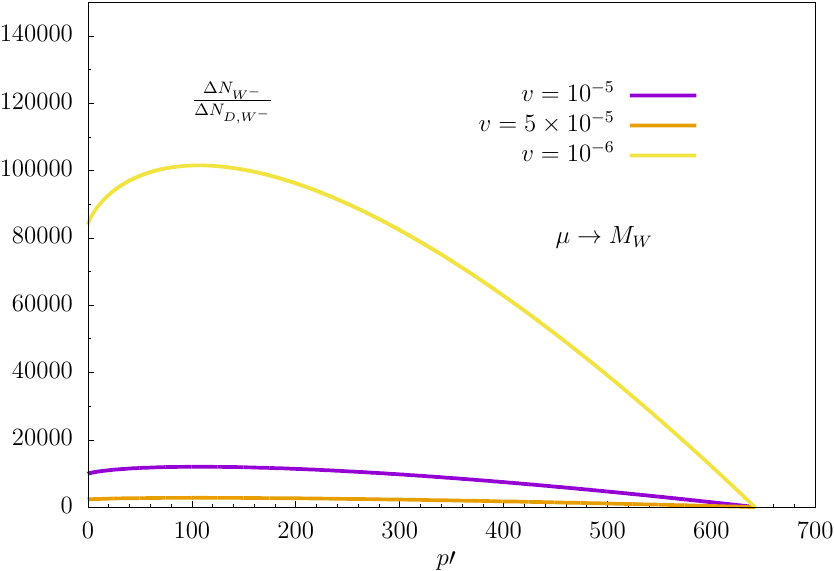}
	\caption{The ratio between the density numbers of produced $W$ bosons in perturbative processes and the density number of decaying $W$ \eqref{nnd} bosons with respect to the neutrino momentum $p\prime$. Here the $W$ boson momentum is $P= \gamma\, M_{W}\, v$ and the electron momentum is $p = \gamma\, m_{\text{el}}\, v$, where $M_{W}=80.4$ GeV is the $W$ boson mass, $m_{\text{el}}=0.511$ MeV and $\gamma=1/\sqrt{1-v^2}$ is the Lorentz factor. The ratio was taken at various relativistic (top figure) and norelativistic (bottom figure) speeds (we take $c=1$), for the case when $\mu=M_W$ in \eqref{nnd}.}
	\label{fig.11}
\end{figure}

Next we turn to the ratio between the production and decay rates of the $W$ boson \eqref{nnd}. Figures \ref{fig.8} and \ref{fig.9} represent the ratio \eqref{nnd} with respect to the electron momentum $p$, while figures \ref{fig.10} and \ref{fig.11} represent the ratio \eqref{nnd} with respect to the neutrino momentum $p\prime$. The ratio has been plotted at various relativistic and nonrelativistic speeds $v$, with respect to $c=1$. In both the electron and neutrino momentum cases, we can see that the ratio between the produced and decaying $W$ bosons is far larger when the speeds of the particles with fixed values for the momenta are smaller, in the case when all logarithms go to 0 (figs. \ref{fig.8} and \ref{fig.10}). At speeds close to the speed of light, the ratio is significantly smaller. The same holds for the case when $\mu \rightarrow M_W$, however, here the ratio decreases also with the varying electron momentum (fig. \ref{fig.9}) and varying neutrino momentum (fig. \ref{fig.11}). The conclusion that can be drawn is that the production of particles with small momenta (speeds) is favoured in the conditions of the Early Universe.

\newpage

\newpage

\begin{figure}[H]
        \centering
	\includegraphics[width=11cm]{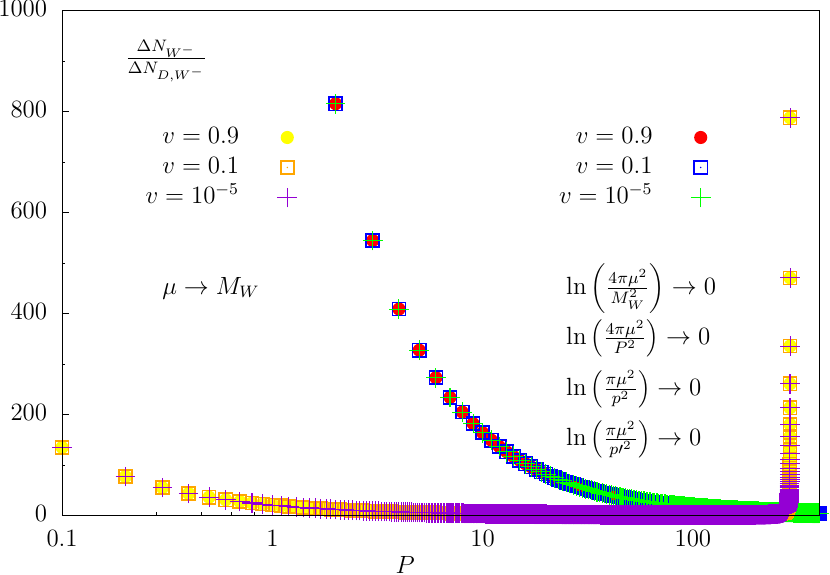}
	\caption{The ratio between the density numbers of produced $W$ bosons in perturbative processes and the density number of decaying $W$ \eqref{nnd} bosons with respect to the $W$ boson momentum $P$. Here the electron momentum is $p= \gamma\, m_{\text{el}}\, v$, where $m_{\text{el}}=0.511$ MeV is the electron mass, the neutrino momentum was taken to be $p\prime = 200$ GeV and $\gamma=1/\sqrt{1-v^2}$ is the Lorentz factor. The ratio was taken at various relativistic ($v=0.9,v=0.1$) and nonrelativistic ($v=10^{-5}$) speeds, since $c=1$. The red, blue and green points represent the ratio when all logarithms in \eqref{nnd} go to 0, while the yellow, orange and purple points represent the case when $\mu \rightarrow M_W$ in \eqref{nnd}. We can see that the values of the ratio do not vary much with respect to the speed $v$, but depend more on the values we choose for the renormalisation mass $\mu$. The x-axis was taken in a logarithmic scale.}
	\label{fig.7}
\end{figure}

In fig. \ref{fig.7} the ratio between the produced $W$ bosons and decaying $W$ bosons is presented in terms of the momentum $P$ of the $W$ boson. The ratio has been plotted at various relativistic and nonrelativistic speeds $v$. We can observe that the ratio becomes important for small $W$ momenta, since the ratio behaves as a $P^{-1}$ function, in the case when all logarithms in \eqref{nnd} go to 0. However, as was the case for the transition rates, when $\mu \rightarrow M_W$, the logarithm dictates the behaviour of the ratio, which now has two peaks: one for $P \rightarrow 0$, and another around the value $P=290$ GeV. It can also be seen that, as opposed to the figures with respect to the electron and neutrino momenta, here the values of the speeds of the other particles do not affect the ratio as much as the value we choose for the renormalisation mass $\mu$. 

\begin{figure}[H]
        \centering
        \begin{subfigure}{0.62\textwidth}
            \includegraphics[width=\textwidth]{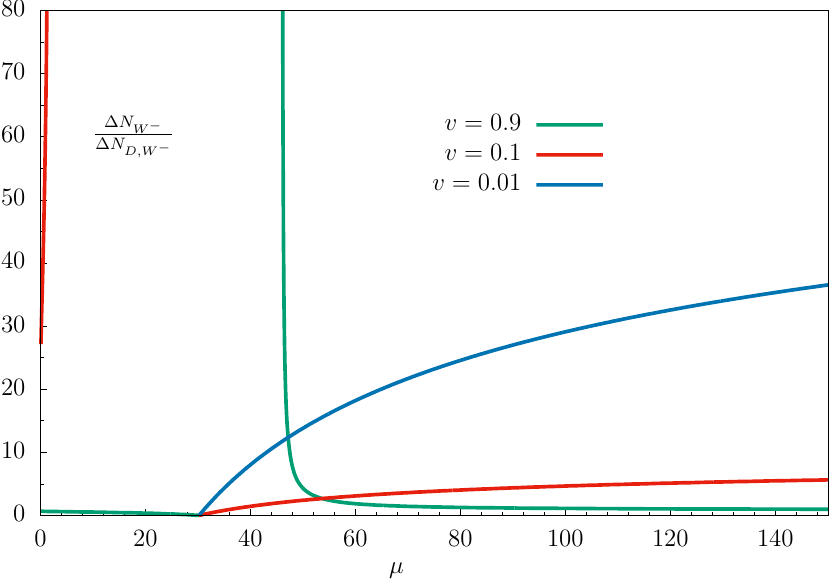}
            \caption{}
            \label{fig.12.sub.1}
        \end{subfigure}
	    \begin{subfigure}{0.37\textwidth}
	        \includegraphics[width=\textwidth]{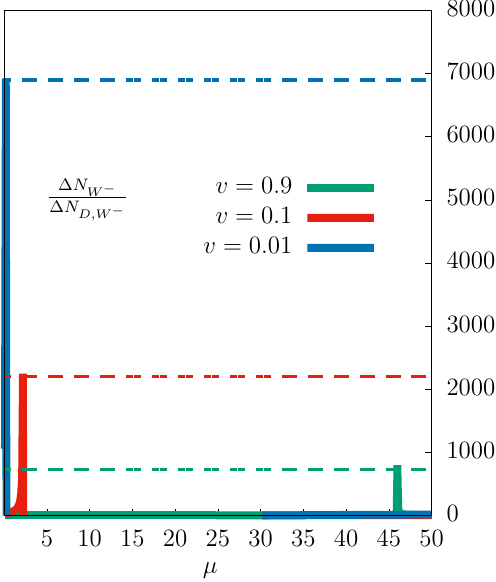}
            \caption{}
            \label{fig.12.sub.2}
	    \end{subfigure}
        \hfil
         \begin{subfigure}{0.8\textwidth}
	        \includegraphics[width=\textwidth]{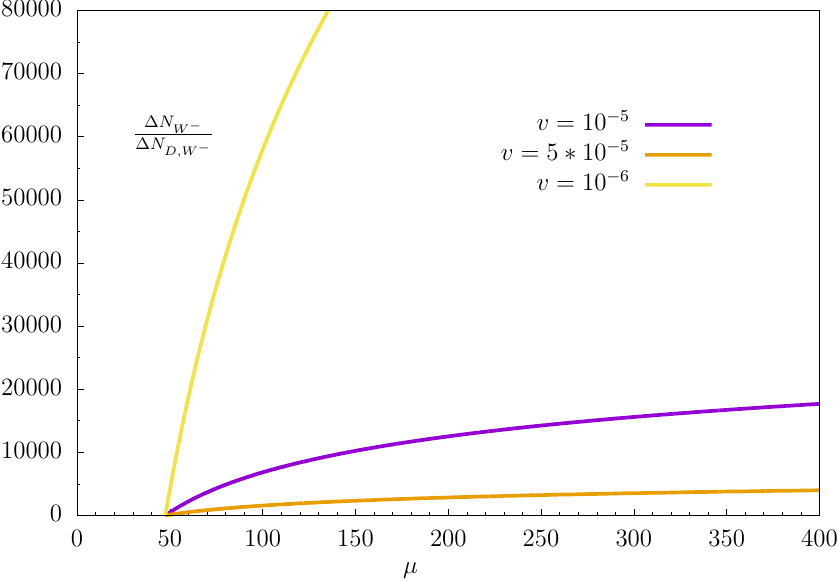}
            \caption{}
            \label{fig.12.sub.3}
	    \end{subfigure}
	\caption{The ratio between the density numbers of produced $W$ bosons in perturbative processes and the density number of decaying $W$ \eqref{nnd} bosons with respect to the renormalisation mass $\mu$. Here the $W$ boson momentum is $P= \gamma\, M_{W}\, v$, the electron momentum is $p = \gamma\, m_{\text{el}} v$ and the neutrino momentum was taken to be $p\prime = 200$ GeV, where $M_{W}=80.4$ GeV is the $W$ boson mass, $m_{\text{el}}=0.511$ MeV and $\gamma=1/\sqrt{1-v^2}$ is the Lorentz factor. The ratio was taken at various relativistic (top figures \ref{fig.12.sub.1} and \ref{fig.12.sub.2}) and nonrelativistic (bottom figure \ref{fig.12.sub.3}) speeds (we take $c=1$). Figure \ref{fig.12.sub.2} is a version of figure \ref{fig.12.sub.1} zoomed in on the interval $\mu \in (0,50 ]$. In figure \ref{fig.12.sub.2} we can see the peaks of the ratio \eqref{nnd}: for example, for $v=0.01$, the ratio has a peak at a value around $7000$, when $\mu =0.2$. At $\mu=2.2$ there is a peak for the ration when $v=0.1$, while at $\mu = 46$ there is a peak for when $v=0.9$.}
	\label{fig.12}
\end{figure}

Figure \ref{fig.12} shows ratio \eqref{nnd} with respect to the renormalisation parameter $\mu$, also at various relativistic and nonrelativistic speeds. In accordance with our previous conclusions, the ratio is higher at nonrelativistic speeds. For relativistic speeds, the ratio seems to peak at certain values of $\mu$ (figs. \ref{fig.12.sub.1} and \ref{fig.12.sub.2}), while for nonrelativistic speeds, the ratio increases as $\mu$ increases to larger values (fig. \ref{fig.12.sub.3}).

In our analysis we have taken into account only the transversal modes of the Proca field in de Sitter spacetime \cite{2}. This choice is justified by the fact that, at earlier times and high temperatures, the mass of the $W$ bosons could be smaller or close to zero, since the mass is related to the expectation values of the Higgs field. Another important step for a better understanding of physics in the Early Universe would require us to establish the variation of the $W$ boson mass with the temperature, via the expansion parameter $\omega$, i.e. a dependence of the form $M_W[T(\omega)]$.

In the end of this section, we must point out that the present density numbers account for the perturbative treatment of the $W$ boson production in the Early Universe, and do not take into account other possible mechanisms of $W$ boson generation. Our results are exclusively the consequence of the large expansion in the early moments of the Universe.

\section{Conclusions}

In this paper we analyze the production of charged $W^{\pm}$ bosons in emission processes by neutrinos and antineutrinos. This phenomenon received little attention in the literature, and our study is the first to take into account neutrino decays as a source of producing massive $W^{\pm}$ bosons.
Our analytical and graphical results prove that such transitions in the first order of electroweak theory, in de Sitter spacetime, can occur only in the large expansion conditions of the Early Universe,
when the background energy was larger than the rest energy of the $W$ bosons. Our graphical analysis also reveals that we obtain the Minkowski limit of neutrino decays, i.e. the probabilities and transition rates become zero when the Hubble constant vanishes $(\omega=0)$. This confirms the fact that the emission of $W$ bosons by neutrinos (antineutrinos), $\nu \rightarrow W^{+} +e^{-}$ ($\widetilde{\nu} \rightarrow W^{-} +e^{+})$, is forbidden in flat spacetime.

Another important result is related to the large expansion limit, when $m/\omega = M_W/\omega =0$, and our equations simplify. In this limit, we applied dimensional regularization and minimal subtraction in order to extract finite results for our total transition rates. Using previous results related to the decay and production of $W^{\pm}$ bosons in de Sitter spacetime, (\cite{44} and \cite{PTEP}), we managed to obtain the density number of $W$ bosons as a function of the particle momenta and renormalisation mass $\mu$. The density number obtained with our perturbative methods is finite in the limit of large expansion from the Early Universe, i.e. $\omega\gg M_W$.

\newpage

\textbf{Acknowledgements}

This work is supported by the European Union - NextGenerationEU through
the grant No. 760079/23.05.2023, funded by the Romanian ministry of
research, innovation and digitalization through Romania’s National
Recovery and Resilience Plan, call no. PNRR-III-C9-2022-I8.

\appendix 

\section*{Appendix A: Free field solutions} \label{AppA}

The positive frequency solution with a well defined momentum $\vec{p}$ and helicity $\sigma$ for the Dirac equation in de Sitter spacetime is \cite{22}:
\begin{equation}\label{solelectron}
  u_{\vec{p},\sigma}(t,\vec{x}) = i\sqrt{\frac{\pi p}{\omega}}\left(\frac{1}{2\pi}\right)^{3/2}
  \begin{pmatrix}
   \frac{1}{2} e^{\pi k/2} H_{\nu_{-}}^{(1)} \big((p/\omega) \, e^{-\omega t}\big) \xi_\sigma (\vec{p}\,) \\
   \sigma e^{-\pi k/2} H_{\nu_{+}}^{(1)} \big( (p/\omega)\, e^{-\omega t}\big) \xi_\sigma (\vec{p}\,)
   \end{pmatrix}
   e^{i\vec{p}\cdot\vec{x}-2\omega t},
\end{equation}
where $H_{\nu}^{(1)}(z)$ are Hankel functions of the first kind and $ \xi_\sigma (\vec{p}\,)$ are the helicity bispinors. The order of the Hankel functions $\nu_{\pm} = 1/2 \pm  i k $ depends on the fermion mass, in this case the mass of the electron, and on the Hubble parameter through $k=m_e/\omega$.

The zero mass Dirac solution in the de Sitter geometry corresponding to the neutrino reads \cite{22}:
\begin{equation}\label{solneutrin}
[{{u^{0}_{\vec{p'}\sigma'}}(x)}]^{\dag} = \left(\frac{-\omega t_c}{2\pi}\right)^{3/2} \Bigg(\left(\frac{1}{2}-\sigma'\right)\xi_{\sigma'}^{\dag}(\vec{p'}),0 \Bigg) e^{-i\vec{p'}\vec{x} + ip't_c},
\end{equation}
where $\vec{p'}$ and $\sigma'$ are the momentum and helicity of the neutrino.

For the Proca field in de Sitter spacetime we have a separation of solutions with respect to the helicity $\lambda$. The $\lambda=\pm 1$ transversal modes retain only the spatial part, because the temporal component vanishes \cite{2}:
\begin{align}\label{procasollambda1}
f_{\vec{P}\lambda=\pm1,\,j}^* (x) = \frac{\sqrt{\pi} e^{-\pi K/2}}{2(2\pi)^{3/2}} \sqrt{-t_{c}} H_{-iK}^{(2)}(-P t_{c}) e^{-i\vec{P}\vec{x}} {\epsilon}^{*}_{j}(\vec{P},\lambda=\pm 1),
\end{align}
where $H_{-iK}^{(1)}(-P t_{c})$ are again Hankel functions of first kind, with their order depending on the $W$ boson mass and Hubble parameter: $K=\sqrt{\left(M_W/\omega\right)^{2}-1/4}$. Furthermore, $\vec{\epsilon}\,(\vec{P},\lambda)$ are the polarization vectors. For $\lambda=\pm 1$ these vectors are orthogonal with the momentum $\vec{P}\cdot\vec{\epsilon}\,\left({P},\lambda=\pm 1 \right)=0$.

For longitudinal Proca modes with $\lambda = 0$ we have a non-vanishing temporal component \cite{2}:
\begin{equation}\label{procalambda0t}
f_{\vec{P}\lambda=0,\,0}(x)=\frac{\sqrt{\pi}e^{-\pi K/2}\omega
P(-t_{c})^{3/2}}{2(2\pi)^{3/2}M_{W}}\,H_{iK}^{(1)}(-Pt_{c})e^{i\vec{P}\cdot\vec{x}},
\end{equation}
while for the spatial component we have:
\begin{align}\label{procalambda0s}
f_{\vec{P}\lambda=0, \,j}(x)=\frac{i\sqrt{\pi}e^{-\pi K/2}\omega P}{2(2\pi)^{3/2}M_{W}}\Bigg[&\Bigl(\frac{1}{2}+iK\Bigl)\frac{\sqrt{-t_{c}}}{P} H_{iK}^{(1)}(-Pt_{c})\nonumber\\
&-(-t_{c})^{3/2} H_{1+iK}^{(1)}(-Pt_{c})\Bigg]
\epsilon_{j}(\vec{P},\lambda)e^{i\vec{P}\cdot\vec{x}}.
\end{align}

The longitudinal Proca modes will be used in \hyperref[AppC]{Appendix C}.

\section*{Appendix B: Bessel functions}\label{AppB}

To transform the Hankel functions and exponentials, which appear in the amplitude formulas, into Bessel functions, we employ the following relations \cite{AS,21}:
\begin{align}
\label{exptok} e^{-z} &= \sqrt{\frac{2 z}{\pi}}K_{1/2}(z), \\
\label{hankeltok} H^{(2),(1)}_{\nu} (z) &= \pm \frac{2 i}{\pi} e^{\pm i\pi\nu/2}K_{\nu}(\pm i z), \\
\label{hankeltoj} H^{(2),(1)}_{\nu} (z) &= \frac{J_{-\nu}(z)-e^{\pm i\pi\nu}J_{\nu}(z)}{\mp i \sin{(\pi\nu)}},
\end{align}
where $J_{\nu}(z)$ are Bessel functions of the first kind and $K_{\nu}(z)$ are modified Bessel functions.

The temporal integrals in the amplitude, obtained using the above formulas, can be solved by making use of the general integral \cite{AS,21}:
\begin{align}\label{appell}
\int_{0}^{\infty} & x^{Q-1} J_{\lambda}(ax) J_{\mu}(bx) K_{\nu}(cx) dx = \nonumber\\
&= \frac{2^{Q-2}a^{\lambda}b^{\mu}c^{-Q-\lambda-\mu}}
{\Gamma(\lambda+1)\Gamma(\mu+1)} \Gamma\left(\frac{Q+\lambda+\mu-\nu}{2}\right) \Gamma\left(\frac{Q+\lambda+\mu+\nu}{2}\right)\nonumber\\
&\times F_{4} \left(\frac{Q+\lambda+\mu-\nu}{2},\frac{Q+\lambda+\mu+\nu}{2},\lambda+1,\mu+1,-\frac{a^{2}}{c^{2}},-\frac{b^{2}}{c^{2}}\right),
\end{align}
where  $F_{4}$ is Appell$^{\prime}$s function.

The Appell $F_4$ function has the following double series expression \cite{AS}:
\begin{align}\label{F4series}
	F_{4}(\alpha,\beta,\gamma,\gamma';x,y) &= \sum_{m=0}^{\infty} \sum_{n=0}^{\infty} \frac{\Gamma(\alpha+m+n)\Gamma{
			(\gamma)}}{\Gamma(\alpha)\Gamma(\gamma+m)} \frac{\Gamma(\beta + m + n)\Gamma(\gamma')}{\Gamma(\beta)\Gamma(\gamma' +n)}\frac{x^{m}y^{n}}{m!n!},
\end{align}
where $\Gamma(z)$ are the Gamma functions.

\section*{Appendix C: Amplitude with longitudinal Proca modes}\label{AppC}

In the case when we use the longitudinal modes for computing the amplitude we have two terms that give contribution:
\begin{align}
\mathcal{A}_{[\nu \rightarrow W^{+} + e^{-}]}^{[\lambda = 0]} &=\frac{ig}{2\sqrt{2}}\int\;d^{4}x\sqrt{-g(x)}\, \bar{u}_{\vec{p}\sigma}(x) \gamma^{\hat{\alpha}}e^{\mu}_{\hat{\alpha}}(1-\gamma^{5}) u_{\vec{p'}\sigma'}^{0}(x)
f^{*}_{\vec{P}\lambda,\,\mu}(x)\nonumber\\
&=\frac{ig}{2\sqrt{2}}\int\;d^{4}x\sqrt{-g(x)}\, \bar{u}_{\vec{p}\sigma}(x) \gamma^{\hat{0}}e^{0}_{\hat{0}}(1-\gamma^{5}) u_{\vec{p'}\sigma'}^{0}(x)
f^{*}_{\vec{P}\lambda,\,0}(x)\nonumber\\
&+\frac{ig}{2\sqrt{2}}\int\;d^{4}x\sqrt{-g(x)}\, \bar{u}_{\vec{p}\sigma}(x)\gamma^{\hat{i}}e^{j}_{\hat{i}}(1-\gamma^{5}) u_{\vec{p'}\sigma'}^{0}(x)
f^{*}_{\vec{P}\lambda,\,j}(x),
\end{align}
where the two terms correspond to the temporal \eqref{procalambda0t} and spatial \eqref{procalambda0s} components of the Proca modes \cite{2}.

In our computations we will split the amplitude into three terms:
\begin{align}\label{a123}
	\mathcal{A}_{[\nu \rightarrow W^{+} + e^{-} ]}^{[\lambda = 0]} = \,\,\,\mathcal{A}^{[\lambda=0](1)}_{[\nu \rightarrow W^{+} + e^{-} ]} + \mathcal{A}^{[\lambda=0](2)}_{[\nu \rightarrow W^{+} + e^{-} ]} + \mathcal{A}^{[\lambda=0](3)}_{[\nu \rightarrow W^{+} + e^{-} ]}.
\end{align}

After solving the spatial integrals and making the substitution $z=e^{-\omega t}/\omega$ we obtain:
\begin{align}
\mathcal{A}^{[\lambda=0](1)}_{[\nu \rightarrow W^{+} + e^{-} ]} &= -\frac{g\pi\sqrt{p} P e^{-\pi K/2} e^{\pi k/2} \delta^{3}(\vec{p}\;'-\vec{p}-\vec{P})}{4\sqrt{2}(2\pi)^{3/2}\left(M_{W}/\omega\right)} \Bigl(\frac{1}{2}-\sigma'\Bigl)\nonumber\\
&\times\int_{0}^{\infty}dz\;z^{2}\, H_{\nu^{+}}^{(2)}(pz) H_{-iK}^{(2)}(Pz) \xi_{\sigma}^{\dag}(\vec{p}\,) \xi_{\sigma'}(\vec{p}\,')e^{ip'z},\\
\mathcal{A}^{[\lambda=0](2)}_{[\nu \rightarrow W^{+} + e^{-} ]} &= \frac{ig\pi\sqrt{p} P e^{-\pi K/2} e^{\pi k/2} \delta^{3}(\vec{p}\;'-\vec{p}-\vec{P})}{4 \sqrt{2} (2\pi)^{3/2}
\left(M_{W}/\omega \right)} \Bigl(\frac{1}{2}-\sigma'\Bigl) \left(\frac{\frac{1}{2}-iK}{P}\right) \nonumber\\
&\times\int_{0}^{\infty}dz\, z\, H_{\nu^{+}}^{(2)}(pz)H_{-iK}^{(2)}(Pz)\xi_{\sigma}^{\dag}(\vec{p}\,)\vec{\sigma}\cdot
\vec{\epsilon}(P,\lambda)\xi_{\sigma'}(\vec{p}\,')e^{ip'z},\\
\mathcal{A}^{[\lambda=0](3)}_{[\nu \rightarrow W^{+} + e^{-} ]} &= \frac{ig\pi\sqrt{p}Pe^{-\pi K/2}e^{\pi k/2}\delta^{3}(\vec{p}\;'-\vec{p}-\vec{P})}{4\sqrt{2}(2\pi)^{3/2}\left(M_{W}/\omega\right)} \Bigl(\frac{1}{2}-\sigma'\Bigl)\nonumber\\
&\times\int_{0}^{\infty}dz\;z^{2}H_{\nu^{+}}^{(2)}(pz)H_{1-iK}^{(2)}(Pz)\xi_{\sigma}^{\dag}(\vec{p}\,)\vec{\sigma}\cdot
\vec{\epsilon}(P,\lambda)\xi_{\sigma'}(\vec{p}\,')e^{ip'z}.
\end{align}

The above terms that contribute to the amplitude can be rewritten after transforming the Hankel functions into Bessel functions using relations from \hyperref[AppB]{Appendix B}. Then the temporal integrals can be solved using formula \eqref{appell}. In the end we obtain the following three contributions to the transition amplitude corresponding to longitudinal Proca modes:

\begin{align}
\label{alambda01} \mathcal{A}^{[\lambda=0](1)}_{[\nu \rightarrow W^{+} + e^{-} ]} &= -\frac{g \sqrt{\pi} \sqrt{pp'} P e^{\pi k/2} \delta^{3}(\vec{p}\;'-\vec{p}-\vec{P})}{ 4 (2\pi)^{3/2} \left(M_{W}/\omega\right) \cosh{(\pi k)}} \Bigl(\frac{1}{2}-\sigma'\Bigl) \nonumber\\ 
&\times\xi_{\sigma}^{\dag}(\vec{p}\,)\xi_{\sigma'}(\vec{p}\,')\left(T_{5}+T_{6}+T_{7}+T_{8}\right),\\
\label{alambda02} \mathcal{A}^{[\lambda=0](2)}_{[\nu \rightarrow W^{+} + e^{-} ]} &= i\frac{g \sqrt{\pi} \sqrt{pp'} e^{\pi k/2} \delta^{3}(\vec{p}\;'-\vec{p}-\vec{P})}{4 (2\pi)^{3/2}\left(M_{W}/\omega \right) 
\cosh(\pi k)} \Bigl(\frac{1}{2}-\sigma'\Bigl) \left( \frac{1}{2}-iK\right) \nonumber\\
&\times\xi_{\sigma}^{\dag}(\vec{p}\,)\vec{\sigma}\cdot
\vec{\epsilon}(P,\lambda)\xi_{\sigma'}(\vec{p}\,')\left(T_{1}+T_{2}+T_{3}+T_{4}\right),  \\
\label{almabda03} \mathcal{A}^{[\lambda=0](3)}_{[\nu \rightarrow W^{+} + e^{-} ]} &= -\frac{g\sqrt{\pi}\sqrt{pp'}Pe^{\pi k/2}\delta^{3}(\vec{p}\;'-\vec{p}-\vec{P})}{4(2\pi)^{3/2}
\left(M_{W}/\omega\right) \cosh(\pi k)} \Bigl(\frac{1}{2}-\sigma'\Bigl)\nonumber\\
&\times\xi_{\sigma}^{\dag}(\vec{p}\,)\vec{\sigma}\cdot
\vec{\epsilon}(P,\lambda)\xi_{\sigma'}(\vec{p}\,')\left(T_{9}+T_{10}+T_{11}+T_{12}\right).
\end{align}

In equation  (\ref{alambda02}), terms $T_{1}, T_{2}, T_{3}$ and $T_{4}$ are the same terms defined for $\lambda=\pm 1$. The rest of the terms have a similar structure to $T_{1}, T_{2}, T_{3},T_{4}$ and are given below:
\begin{align}
T_{5} &=\frac{ie^{-\pi
k}\sqrt{2}^{3}p^{\frac{1}{2}+ik}p'^{-\frac{1}{2}}(iP)^{-\frac{7}{2}-ik}}{\Gamma\left(\frac{3}{2}+ik\right)\Gamma\left(\frac{1}{2}\right)}\\\nonumber
&\times\Gamma\left(\frac{7}{4}+\frac{i(k+K)}{2}\right)\Gamma\left(\frac{7}{4}+\frac{i(k-K)}{2}\right)\\\nonumber
&\times F_{4}\Bigg(\frac{7}{4}+\frac{i(k+K)}{2},\frac{7}{4}+\frac{i(k-K)}{2},\frac{3}{2}+ik,\frac{1}{2},\frac{p^2}{P^2},\frac{p'^2}{P^2}\Bigg),
\end{align}

\begin{align}
T_{6}&=-\frac{e^{-\pi
k}\sqrt{2}^{3}p^{\frac{1}{2}+ik}p'^{-\frac{1}{2}}(iP)^{-\frac{9}{2}-ik}}{\Gamma\left(\frac{3}{2}+ik\right)\Gamma\left(\frac{3}{2}\right)}\\\nonumber
&\times
\Gamma\left(\frac{9}{4}+\frac{i(k+K)}{2}\right)\Gamma\left(\frac{9}{4}+\frac{i(k-K)}{2}\right)\\\nonumber
&\times F_{4}\Bigg(\frac{9}{4}+\frac{i(k+K)}{2},\frac{9}{4}+\frac{i(k-K)}{2},\frac{3}{2}+ik,\frac{3}{2},\frac{p^2}{P^2},\frac{p'^2}{P^2}\Bigg),
\end{align}

\begin{align}
T_{7}&=-\frac{\sqrt{2}^{3}p^{-\frac{1}{2}-ik}p'^{-\frac{1}{2}}(iP)^{-\frac{5}{2}+ik}}{\Gamma\left(\frac{1}{2}-ik\right)\Gamma\left(\frac{1}{2}\right)}\\\nonumber
&\times
\Gamma\left(\frac{5}{4}-\frac{i(k+K)}{2}\right)\Gamma\left(\frac{5}{4}+\frac{i(K-k)}{2}\right)\\\nonumber
&\times F_{4}\Bigg(\frac{5}{4}-\frac{i(k+K)}{2},\frac{5}{4}+\frac{i(K-k)}{2},\frac{1}{2}-ik,\frac{1}{2},\frac{p^2}{P^2},\frac{p'^2}{P^2}\Bigg),
\end{align}

\begin{align}
T_{8}&=-\frac{i\sqrt{2}^{3}p^{-\frac{1}{2}-ik}p'^{\frac{1}{2}}(iP)^{-\frac{7}{2}+ik}}{\Gamma\left(\frac{1}{2}-ik\right)\Gamma\left(\frac{3}{2}\right)}\\\nonumber
&\times
\Gamma\left(\frac{7}{4}-\frac{i(k+K)}{2}\right)\Gamma\left(\frac{7}{4}+\frac{i(K-k)}{2}\right)\\\nonumber
&\times F_{4}\Bigg(\frac{7}{4}-\frac{i(k+K)}{2},\frac{7}{4}+\frac{i(K-k)}{2},\frac{1}{2}-ik,\frac{3}{2},\frac{p^2}{P^2},\frac{p'^2}{P^2}\Bigg),
\end{align}

\begin{align}
T_{9} &=\frac{ie^{-\pi
k}\sqrt{2}^{3}p^{\frac{1}{2}+ik}p'^{-\frac{1}{2}}(iP)^{-\frac{7}{2}-ik}}{\Gamma\left(\frac{3}{2}+ik\right)\Gamma\left(\frac{1}{2}\right)}\\\nonumber
&\times
\Gamma\left(\frac{5}{4}+\frac{i(k+K)}{2}\right)\Gamma\left(\frac{9}{4}+\frac{i(k-K)}{2}\right)\\\nonumber
&\times F_{4}\Bigg(\frac{5}{4}+\frac{i(k+K)}{2},\frac{9}{4}+\frac{i(k-K)}{2},\frac{3}{2}+ik,\frac{1}{2},\frac{p^2}{P^2},\frac{p'^2}{P^2}\Bigg),
\end{align}

\begin{align}
T_{10}=&-\frac{e^{-\pi
k}\sqrt{2}^{3}p^{\frac{1}{2}+ik}p'^{\frac{1}{2}}(iP)^{-\frac{9}{2}-ik}}{\Gamma\left(\frac{3}{2}+ik\right)\Gamma\left(\frac{3}{2}\right)}\\\nonumber
&\times
\Gamma\left(\frac{7}{4}+\frac{i(k+K)}{2}\right)\Gamma\left(\frac{11}{4}+\frac{i(k-K)}{2}\right)\\\nonumber
&\times F_{4}\Bigg(\frac{7}{4}+\frac{i(k+K)}{2},\frac{11}{4}+\frac{i(k-K)}{2},\frac{3}{2}+ik,\frac{3}{2},\frac{p^2}{P^2},\frac{p'^2}{P^2}\Bigg),
\end{align}

\begin{align}
T_{11}&=-\frac{\sqrt{2}^{3}p^{-\frac{1}{2}-ik}p'^{-\frac{1}{2}}(iP)^{-\frac{5}{2}+ik}}{\Gamma\left(\frac{1}{2}-ik\right)\Gamma\left(\frac{1}{2}\right)}\\\nonumber
&\times
\Gamma\left(\frac{7}{4}-\frac{i(k+K)}{2}\right)\Gamma\left(\frac{3}{4}+\frac{i(K-k)}{2}\right)\\\nonumber
&\times F_{4}\Bigg(\frac{7}{4}-\frac{i(k+K)}{2},\frac{3}{4}+\frac{i(K-k)}{2},\frac{1}{2}-ik,\frac{1}{2},\frac{p^2}{P^2},\frac{p'^2}{P^2}\Bigg),
\end{align}

\begin{align}
T_{12}&=-\frac{i\sqrt{2}^{3}p^{-\frac{1}{2}-ik}p'^{\frac{1}{2}}(iP)^{-\frac{7}{2}+ik}}{\Gamma\left(\frac{1}{2}-ik\right)\Gamma\left(\frac{3}{2}\right)}\\\nonumber
&\times
\Gamma\left(\frac{9}{4}-\frac{i(k+K)}{2}\right)\Gamma\left(\frac{5}{4}+\frac{i(K-k)}{2}\right)\\\nonumber
&\times
F_{4}\Bigg(\frac{9}{4}-\frac{i(k+K)}{2},\frac{5}{4}+\frac{i(K-k)}{2},\frac{1}{2}-ik,\frac{3}{2},\frac{p^2}{P^2},\frac{p'^2}{P^2}\Bigg).
\end{align}

We can observe that the amplitude for longitudinal Proca modes has a factor of $\omega/M_W$ via \eqref{alambda01}, \eqref{alambda02} and \eqref{almabda03}. The factor of $\omega$ makes it difficult to analyze the transition rate corresponding to this amplitude in the large expansion limit, when $\omega \rightarrow \infty$. That's why we restricted ourselves only to transversal modes when we computed the transition rate of the neutrino decay.


\begin{thebibliography}{99}
\bibitem{1}
C. W. Misner, K. S. Thorne and J. A. Wheleer, {\em Gravitation}
(W. H. Freeman and Company New York, 1973).
\bibitem{PR1}
A. Proca, {\em J. Phys. Radium} \textbf{7}, 347–353 (1936).
\bibitem{PR2}
A. Proca, {\em C. R. Acad. Sci. Paris}, \textbf{202}, 1366 (1936);\,
A. Proca, {\em C. R. Acad. Sci. Paris} \textbf{202}, 1490 (1936). 
\bibitem{PR3}
A. Proca, {\em  J. Phys. Radium} \textbf{9}, 61 (1938).
\bibitem{2}
Ion I. Cot\u{a}escu, {\em Gen.Rel.Grav.} \textbf{42},861-876,2010.
\bibitem{w1}
S. Weinberg, The First Three Minutes: A Modern View of the Origin of the Universe (Basic Books,New York, 1977).
\bibitem{w2}
S. Weinberg, {\em Phys. Scr.} \textbf{21}, 773 (1979).
\bibitem{3}
S. Weinberg, {\em Phys. Rev. Lett.} \textbf{19}, 1264 (1967).
\bibitem{4}
S. Weinberg, {\em Phys. Rev. Lett.} \textbf{27}, 1688 (1971).
\bibitem{5}
S. Weinberg, {\em Phys. Rev. D} \textbf{5}, 1412 (1972).
\bibitem{6}
S. Weinberg, {\em Phys. Rev. D} \textbf{7}, 1068 (1973).
\bibitem{7}
S. Weinberg, {\em Phys. Rev. D} \textbf{8}, 605 (1973).
\bibitem{8}
S. Weinberg, {\em Rev. Mod. Phys.} \textbf{46}, 255 (1974).
\bibitem{9}
S. L. Glasshow and S. Weinberg, {\em Phys. Rev. D} \textbf{15}, 1958 (1977).
\bibitem{10}
J. D. Bjorken, K. Lane and S. Weinberg, {\em Phys. Rev. D} \textbf{5}, 1474 (1977).
\bibitem{11}
B. W. Lee and S. Weinberg, {\em Phys. Rev. D} \textbf{38}, 1237 (1977).
\bibitem{cr}
C. Rubbia, {\em Rev. Mod. Phys.} \textbf{57}, 699 (1985).
\bibitem{12}
S. Weinberg, {\em The Quantum Theory of Fields}  (Cambridge University Press, Cambridge, 1995).
\bibitem{15}
J. Lankinen and I. Vilja, {\em Phys. Rev. D} \textbf{96}, 105026-1 (2017).
\bibitem{17}
N. D. Birrel and P. C. W. Davies,  {\em Quantum Fields in Curved Space} (Cambridge University Press, Cambridge 1982).
\bibitem{18}
N. D. Birrel, P. C. W. Davies and L. H. Ford, {\em J. Phys. A} \textbf{13}, 961 (1980).
\bibitem{19}
S. Drell and J. D. Bjorken, {\em Relativistic Quantum Fields} (Mc Graw-Hill Book Co., New York 1965).
\bibitem{20}
L. Landau and E. M. Lifsit, {\em Theorie Quantique Relativiste} (Mir Moscou 1972).
\bibitem{AS}
M. Abramowitz and I. A. Stegun, {\em Handbook of Mathematical Functions} (Dover, New York, 1964).
\bibitem{21}
I. S. Gradshteyn and I. M. Ryzhik {\em Table of integrals, series and products} (Academic Press, 2007).
\bibitem{PT}
T. Prokopec, N. C. Tsamis and R. P. Woodard, {\em AnnalsPhys.} \textbf{323}, 1324-1360,(2008).
\bibitem{WT}
N. C. Tsamis and R. P. Woodard, {\em J.Math.Phys.} \textbf{48}, 052306, (2007).
\bibitem{PC}
J. F. Koksma, T. Prokopec, {\em Class.Quant.Grav.} \textbf{26}, 125003, (2009).
\bibitem{CRR}
P. Candelas and D. J . Raine, {\em Phys. Rev. D} \textbf{12}, 965, (1975).
\bibitem{COT}
Ion I. Cot\u{a}escu, {\em Eur. Phys. J. C} \textbf{78:769}, (2018).
\bibitem{WF}
M.E. Fisher and K.G. Wilson, {\em Phys. Rev. Lett.} \textbf{28}, 240 (1972);
K.G. Wilson, {\em Phys. Rev. D} \textbf{7}, 2911 (1973).
\bibitem{GHV}
G.’t Hooft and M. Veltman, {\em Nucl. Phys. B} \textbf{44}, 189 (1972).
\bibitem{IT}
C.G. Bollini, J.J. Giambiagi, {\em Nuovo Cimento B} \textbf{12}, 20 (1972).
\bibitem{GT}
G.’t Hooft, {\em Nucl. Phys. B} \textbf{61}, 455 (1973).
\bibitem{PV}
W. Pauli and F. Villars, {\em Rev. Mod. Phys.} \textbf{21}, 434 (1949).
\bibitem{HGF}
H. Kleinert and V. Schulte-Frohlinde, {\em Critical properties of $\phi^4$-theories} (World Scientific 2001).
\bibitem{22}
Ion I. Cot\u{a}escu, {\em Phys. Rev. D} \textbf{65}, 084008 (2002).
\bibitem{LL}
J. Lankinen and I. Vilja, {\em Phys. Rev. D} \textbf{96}, 105026-1 (2017)
\bibitem{LL1}
J. Lankinen, J. Malmi and I. Vilja, {\em Eur. Phys. J. C} \textbf{80:502}, (2020).
\bibitem{CML}
B. Carter and R. G. McLenaghan, {\em Phys. Rev. D}, {\bf 19} (1979) 1093.
\bibitem{23}
Crucean Cosmin, {\em Phys. Rev. D} \textbf{85}, 084036 (2012).
\bibitem{rc}
C. Crucean and R. Racoceanu, {\em Int. J. Mod. Phys. A} \textbf{23}, (2008).
\bibitem{24}
Ion I. Cot\u{a}escu, C. Crucean, {\em Phys. Rev. D} \textbf{87}, 044016 (2013).
\bibitem{25}
Ion I. Cot\u{a}escu, C. Crucean, {\em Progress of Theor. Phys.} \textbf{124}, 1051 (2010).
\bibitem{26}
C. Crucean and M. A. B\u{a}loi {\em Phys. Rev. D} \textbf{93}, 044070 (2016).
\bibitem{27}
C. Crucean, {\em Mod. Phys. Lett. A} \textbf{22}, 2573 (2007).
\bibitem{28}
C. Crucean and M. A. B\u aloi, {\em Int. J. Mod. Phys. A} \textbf{30}, 1550088 (2015).
\bibitem{29}
Ion I. Cot\u{a}escu, R. Racoceanu, Radu and C. Crucean, {\em Mod. Phys. Lett. A} \textbf{21}, 1313 (2006).
\bibitem{30}
Ion I. Cot\u{a}escu, C. Crucean, {\em Int. J. Mod. Phys. A} \textbf{23}, 3707 (2008).
\bibitem{31}
M. A. B\u{a}loi, {\em Mod. Phys. Lett. A} \textbf{29}, 1450138 (2014).
\bibitem{b1}
M. A. B\u{a}loi,{\em Int. J. Mod. Phys. A} \textbf{31}, 1650081 (2016).
\bibitem{b2}
M. A. B\u{a}loi, C. Crucean and D. Popescu, {\em Eur. Phys. J. C} \textbf{78:398}, (2018).
\bibitem{cc}
C. Crucean, {\em Eur. Phys. J. C} \textbf{79:483}, (2019).
\bibitem{32}
E. Schr\" odinger, {\em Physica} \textbf{6}, 899 (1939).
\bibitem{33}
L. Parker, {\em Phys. Rev. Lett.} \textbf{21}, 562 (1968).
\bibitem{34}
L. Parker, {\em Phys. Rev.} \textbf{183}, 1057 (1969).
\bibitem{35}
L. Parker, {\em Phys. Rev. D} \textbf{3}, 346 (1971).
\bibitem{gar}
J. Garriga, {\em Phys. Rev. D} \textbf{49}, 6327 (1994).
\bibitem{cpc}
Ion I. Cot\u{a}escu, D. Popescu, {\em Chinese Phys. C} \textbf{44}, (2020).
\bibitem{36}
G Steigman, {\em Nucl. Phys. B }, \textbf{252}, 73 (1985).
\bibitem{rat}
W. N. Cottingham, D. A. Greenwood, {\em An introduction to standard model of particle physics}, (Cambridge University Press, Cambridge 2007).
\bibitem{rfv}
Ion I. Cot\u{a}escu, {\em Eur. Phys. J. C} \textbf{79:696}, (2019).
\bibitem{rfv1}
Ion I. Cot\u{a}escu, {\em Eur. Phys. J. C} \textbf{80:535}, (2020).
\bibitem{38}
B. Allen and T. Jacobson, {\em Commun. Math. Phys.} \textbf{103}, 669-692 (1986).
\bibitem{40}
T. Prokopec, O. Törnkvist, R.P. Woodard, {\em Phys. Rev. Lett.} \textbf{89}, 101301 (2002).
\bibitem{41}
T. Prokopec, O. Törnkvist, R.P. Woodard, {\em Annals of Physics} \textbf{303(2)}, 251-274 (2002).
\bibitem{43}
D. Dumitrele ; M. A. B\u aloi; C. Crucean,  {\em Eur. Phys. J. C} \textbf{83:738}, 2023.
\bibitem{44}
C. Crucean ; A. D. Fodor, {\em Eur. Phys. J. C} \textbf{83:929}, 2023.
\bibitem{45}
C. Crucean, D. Dumitrele {\em Eur. Phys. J. C} \textbf{84:855}, 2024.
\bibitem{46}
M. A. B\u aloi, C. Crucean, {\em Int.J.Mod. Phys. A} \textbf{32}, 1750208, (2017).
\bibitem{PTEP}
A. D. Fodor, C. Crucean, {\em Progress of Theor. and  Phys.} \textbf{053E02}, (2025).
\end{thebibliography}
\end{document}